\documentclass[a4paper,10pt]{article}

\pdfoutput=1
\usepackage{jheppub1}
\usepackage{array}
\usepackage{multirow}
\usepackage{amsmath}
\usepackage{makecell}

\usepackage{cancel}
\usepackage{amsmath,amssymb} 
\usepackage{graphicx} 
\usepackage[dvipsnames,x11names]{xcolor} 

\usepackage{hyperref}
\usepackage{cleveref}
\usepackage{float}
\usepackage{subcaption}
\usepackage{comment}

\makeatletter
\newcommand{\footnoteref}[1]{\protected@xdef\@thefnmark{\ref{#1}}\@footnotemark}
\makeatother

\begin{document}

	\preprint{}
	%\title{Comparing two entropy constructions for higher curvature black holes}
    \title{A comparison of two constructions for dynamical corrections to Wald entropy}
	
	\author[a]{Sayantani Bhattacharyya}
    \affiliation[a]{School of Mathematics and Maxwell Institute for Mathematical Sciences, University of Edinburgh,
	Peter Guthrie Tait Road, Edinburgh EH9 3FD, United Kingdom}

    \author[b]{, Parthajit Biswas}
    \affiliation[b]{Saha Institute of Nuclear Physics, 1/AF Bidhannagar, Kolkata 700064, India}
    
    \author[c]{, and Nilay Kundu}
    \affiliation[c]{Department of Physics
	Indian Institute of Technology Kanpur, Kalyanpur, Kanpur 208016, India}
    
	\emailAdd{ sbhatta5@ed.ac.uk, parthajitbiswas8@gmail.com, nilayhep@iitk.ac.in}

\abstract{In this work, we analyze the differences and similarities between two recent constructions, which are distinct in their methodologies for extending the Wald entropy of stationary black holes to non-stationary situations in general higher-derivative gravity. One of them, denoted by $S_\text{Wall}$, is constructed by exploiting the boost symmetry of the near-horizon geometry, whereas the other, denoted by $S_\text{dyn}$, is obtained from a covariant phase-space analysis based on the Wald-Iyer Noether charge formalism. While $S_\text{dyn}$ is, by construction, defined only for linearized fluctuations around a stationary black hole solution, $S_\text{Wall}$ does not require such a linearization for its construction. Although the linearization is necessary to interpret $S_\text{Wall}$ as a well-defined notion of entropy, the construction itself naturally contains terms that are higher order in the dynamical fluctuations. By comparing the technical structures underlying the two constructions, we clarify the fundamental differences between the methods on which they are based. We demonstrate that while the construction of $S_\text{dyn}$ given the $S_\text{Wall}$ is straightforward, the converse is more subtle. We develop an algorithm to obtain a local expression for $S_\text{Wall}$ from a known expression for $S_\text{dyn}$ in a generic diffeomorphism-invariant theory of gravity,  provided certain technical conditions are satisfied. We justify our analytical findings with explicit demonstrations in a particular case: the Riemann-squared example of the higher-derivative theory of gravity.}

%{In this note, we compare two recent constructions of black hole entropy in higher-curvature gravity, namely the {\it Wall entropy} ($S_{\rm Wall}$) and the {\it dynamical entropy} ($S_{\rm dyn}$), which are based on two different approximation schemes, and clarify the relationship between them. While the construction of $S_{\rm dyn}$ from $S_{\rm Wall}$ is straightforward, the converse is more subtle. We show how to construct $S_{\rm Wall}$ from a given expression for $S_{\rm dyn}$.}

\maketitle

\section{Introduction}\label{intro}
In the two-derivative theory of gravity, namely Einstein's general relativity (GR), black hole solutions satisfy the three laws of thermodynamics upon identifying appropriate black hole quantities with the corresponding thermodynamic variables \cite{Bekenstein:1973ur,Bardeen:1973gs,Bekenstein:1974ax,Hawking:1975vcx,Hawking:1971tu,Hawking:1971vc}. However, Einstein's GR is believed to be only a low-energy effective description of gravity, and any ultraviolet-complete theory is expected to generate higher-curvature corrections to the Einstein-Hilbert action. Black hole solutions continue to exist in the presence of such higher-derivative corrections, and one expects them to obey the laws of black hole thermodynamics, at least in the regime where the higher-derivative terms can be treated perturbatively.

Understanding the thermodynamic laws of black holes in higher-derivative theories of gravity has therefore been an active area of research for several decades. In \cite{Wald:1993nt,Iyer:1994ys}, using a Noether-charge construction, Wald and collaborators proposed an entropy functional, now known as the {\it Wald entropy} $(S_{\rm Wald})$, and showed that it satisfies the first law of black hole thermodynamics. In this construction, the higher-curvature terms need not be treated as perturbative corrections to the Einstein-Hilbert term, implying that the formalism is applicable beyond the regime of the effective field theory expansion. 

Investigations of the zeroth law in higher-curvature theories of gravity have likewise been carried out in the literature. In particular, the validity of the zeroth law has been analyzed in several specific higher-curvature models in \cite{Ghosh:2020dkk,Sarkar:2012wy,Sang:2021rla,Dey:2021rke}. More generally, proofs of the zeroth law have been obtained under additional assumptions, such as {\it staticity} or {\it stationarity together with axisymmetry} \cite{Racz:1995nh}. In the works \cite{Bhattacharyya:2022nqa,Davies:2024fut,Biswas:2025obu}, the authors established a proof of the zeroth law for arbitrary higher-curvature theories of gravity within the framework of effective field theory. The proof requires only that the event horizon be a Killing horizon, and remains valid even in the presence of non-minimally coupled matter fields, including scalar, gauge, and Proca fields.

It is the second law that still has not been proved completely satisfactorily for arbitrary higher derivative theories of gravity. It is well known that the {\it Wald entropy} $(S_{\rm Wald})$ does not, in general, satisfy the second law of black hole thermodynamics \cite{Jacobson:1993xs,Jacobson:1993vj,Jacobson:1995uq}. Consequently, substantial effort has been devoted to understanding the second law in higher-curvature theories of gravity, primarily in the context of specific higher-derivative theories \cite{Sarkar:2013swa,Bhattacharjee:2015yaa,Bhattacharyya:2016xfs,Fairoos:2018pee,Sarkar:2019xfd,Chatterjee:2025jur}. In a major development, in the works \cite{Wall:2015raa,Bhattacharyya:2021jhr}, the authors constructed an entropy functional (denoted by $S_{\rm Wall}$) that satisfies the second law for linearized perturbations about stationary black hole solutions in arbitrary higher-derivative theories of gravity. The works \cite{Wall:2015raa,Bhattacharyya:2021jhr} subsequently stimulated a flurry of activity in black hole thermodynamics \cite{Bhattacharya:2019qal,Bhattacharyya:2022njk,Biswas:2022grc,Chandranathan:2022pfx,Dhivakar:2023mai,Deo:2023vvb,Bhattacharyya:2024xpu,Wall:2024lbd,Kar:2024dqk,Yan:2024gbz,Davies:2022xdq,Davies:2023qaa,Hollands:2022fkn,Hollands:2022ajj,Shajiee:2025cxl}.

More recently, the authors of \cite{Hollands:2024vbe} introduced a dynamical entropy, denoted by $S_{\rm dyn}$, for stationary black holes perturbed by small dynamical fluctuations. In contrast to the Wald entropy, $S_{\rm Wald}$, which is defined only on the bifurcation surface $\mathcal{B}$, $S_{\rm dyn}$ is well defined on an arbitrary cross-section of the event horizon. Employing the physical process version of the first law \cite{Wald:1995yp,Gao:2001ut}, they demonstrated that this entropy satisfies the second law for the same small amplitude perturbations about stationary black hole solutions in arbitrary higher-derivative theories of gravity\footnote{Several subsequent works have further explored various aspects and applications of the dynamical entropy $S_{\rm dyn}$ \cite{Visser:2024pwz,Visser:2025jnf,Furugori:2025pmn,Shajiee:2026coz,Fu:2026ddy}. For recent developments on black hole entropy beyond linearized perturbation, see \cite{Ashtekar:2026jdz,Fu:2026itf}.}. The physical process version of the first law also plays an important role in the construction of the Wall entropy (see Appendix~\ref{App:PPF} for further discussion). However, the arguments used to arrive at the formula for $S_{\rm Wall}$ is quite different than that of $S_{\rm dyn}$, though both of them use the same component of the equations of motion (both of its indices are projected along the null generator of the horizon)  as the starting point.

In this note, we compare these two constructions of black hole entropy and clarify the precise manner in which they differ from one another. To this end, we first present a brief and schematic outline of the two approaches in order to emphasize their essential distinction. A more detailed discussion will be provided in the subsequent sections. Let us denote the particular component of the equations of motion as $E_{vv}$ where $v$ is the affine parameter of the generators of the horizon. In \cite{Hollands:2024vbe}, the authors have argued that at linear order around any stationary black hole, $v\,E_{vv}$ will satisfy the following identity
\begin{equation}\label{eq:sdynfd}
\int_{\Sigma_v}v\,\delta E_{vv}\sim \partial_v\delta S_{\rm dyn}(v)\,,
\end{equation}
for some $S_{\rm dyn}$ identified with black hole entropy (see eqn.\, \eqref{iden6}), here, $\Sigma_v $ is any constant $v$ slice of the horizon and $\delta$ denotes arbitrary linearized fluctuation around a stationary black hole solution. Now suppose the dynamics was initiated because of some infalling matter field with stress tensor $\delta T_{ab}$. It follows that
\begin{equation}
\partial_v\delta S_{\rm dyn}(v) = \int_{\Sigma_v} v\,\delta T_{vv}\,.
\end{equation}
So if the null energy condition ( $\delta T_{vv}>0$) holds, it implies $\delta S_{\rm dyn} (v_f) - \delta S_{\rm dyn} (v_i)  >0$,  {\it i.e.}, the difference in entropy between the final $v_f$ and initial $v_i$ is always positive.\\\\
On the other hand, in \cite{Bhattacharyya:2021jhr}, the same component $E_{vv}$ was shown to follow a different identity
\begin{equation}\label{eq:swlf}
\int_{\Sigma_v}E_{vv}\sim - \partial_v^2 S_{\rm Wall}\,,
\end{equation}
up to terms with more than one positive boost-weight factor (see subsection \S\ref{subsec:Walle} for details). These terms are not small unless we impose additional assumptions. Here, $S_{\rm Wall}$ is identified with the black hole entropy (see eqn.\,\eqref{eq:Wallfnl}). Once we use the approximation of linearized fluctuations around a stationary black hole solution, the extra terms neglected in \eqref{eq:swlf} become of order $(\delta g)^2$ or higher. The main difference from the dynamical entropy construction, however, is that $S_{\rm Wall}$ generally contains terms of higher order in $(\delta g)$. Now, once again, using the null energy condition $\delta T_{vv}>0$, we obtain
\begin{equation}
\int_{\Sigma_v}\delta T_{vv}=\int_{\Sigma_v}\delta E_{vv}\sim - \partial_v^2 \delta S_{\rm Wall}\,.
\end{equation}

In this case the null energy condition not only states that the entropy always increases ({\it i.e.}, the second law), it also says that the rate of entropy increase is monotonically decreasing till it reaches the final equilibrium, where the rate of increase vanishes. This makes the equilibrium to be the state with maximum entropy.

In fact it has also been shown in  \cite{Hollands:2024vbe} that $S_{\rm dyn}$ is not exactly equal to $S_{\rm Wall}$ but rather they satisfy a  differential relation
\begin{equation}\label{c0}
S_{\rm dyn} = (1- v\partial_v)S_{\rm Wall}\,.
\end{equation}
Here $v$ is the affine parameter along the null generators of the horizon. Both $S_{\rm dyn}$ and $S_{\rm Wall}$ reduce to the Wald entropy \cite{Iyer:1994ys} in the stationary limit. In dynamical situations, however, they differ from the Wald entropy by JKM ambiguity terms. In other words, the constructions of $S_{\rm dyn}$ and $S_{\rm Wall}$ correspond to two different prescriptions for fixing the JKM ambiguity.\\\\
In this note, we explore the relationship between these two approaches, namely the one developed in \cite{Hollands:2024vbe} and the other developed in \cite{Wall:2015raa, Bhattacharyya:2021jhr}. In particular, we would like to see whether the existence of $S_{\rm dyn}$ is enough to construct $S_{\rm Wall}$. This seems to be an interesting question since the physical content of $S_{\rm dyn}$ and $S_{\rm Wall}$ is a bit different as explained above. Our main results may be summarized as follows. 
\begin{itemize}
\item The construction of $S_{\rm Wall}$ uses a different kind of approximation scheme than that of the $S_{\rm dyn}$. As a result there exist few terms that are ignored in the construction of $S_{\rm dyn}$, but are taken into account in the construction of $S_{\rm Wall}$.  However, the physical situation where both of them are applied, namely a stationary black hole with linearized dynamical fluctuation, the existence of one of them is enough to construct the other via the differential relation \eqref{c0}.

\item Following the argument in \cite{Hollands:2024vbe}, one could easily see that for for any small-amplitude dynamical perturbation, $S_{\rm dyn}$ satisfies the constraint
\begin{equation}\label{c1}
\partial_v\left(\delta S_{\rm dyn}\right) \vert_{v=0}= 0\,,
\end{equation}
here, $v=0$ denotes the location of the bifurcation surface where the Killing vector of the stationary background vanishes. Since the construction of dynamical entropy requires the split of the metric into stationary background and fluctuation, the surface $v=0$ plays a special role and the coordinate $v$ appears explicitly in the expression of $S_{\rm dyn}$. On the other hand, in the construction of $S_{\rm Wall}$ \cite{Wall:2015raa,Bhattacharyya:2021jhr}, such a split is not required, and the final expression of the entropy does not have any explicit coordinate dependence.

However in \cite{Bhattacharyya:2021jhr}, the authors derived a relation between the relevant components of the Noether charge ${\bf Q}$ and the presymplectic potential $\Theta$ evaluated on the null hypersurface (eqn. \eqref{eq:2105}). This relation played a crucial role in establishing the existence of $S_{\rm Wall}$.

In this note, we show that the condition \eqref{c1} is, in fact, a special case of the relation \eqref{eq:2105}, evaluated on a split geometry at the surface $v=0$.

\item Finally, we show that given a $S_{\rm dyn}$ satisfying \eqref{c1}, it is always possible to construct $S_{\rm Wall}$ locally for arbitrary linearized perturbations around any stationary geometry. But the $S_{\rm Wall}$ thus constructed generally differs from the $S_{\rm Wall}$ constructed in \cite{Bhattacharyya:2021jhr} by terms that are of higher than linear order in amplitude expansion and are therefore negligible in the framework used in \cite{Hollands:2024vbe}, but not negligible under the approximation scheme used in \cite{Bhattacharyya:2021jhr}.
\end{itemize}
The remainder of this note is organized as follows. In section\,\S\ref{sec:comparison}, we review the constructions of $S_{\rm Wall}$ and $S_{\rm dyn}$ and discuss the relationship between them. In section \S\ref{sec:Swall}, we demonstrate how, given an expression for 
$S_{\rm dyn}$, 
one can construct the corresponding $S_{\rm Wall}$. Section\,\S\ref{sec:example} contains an explicit example illustrating our construction in a Riemann-squared theory. In section\,\S\ref{sec:disc}, we discuss the implications of our results and outline several directions for future work. Additional notational conventions and technical details are provided in the various appendices.

%\section{A comparative analysis of the constructions used in \cite{Wall:2015raa,Bhattacharyya:2021jhr} and \cite{Hollands:2024vbe}}\label{sec:comparison}
\section{A comparative analysis of the constructions of $S_{\rm dyn}$ and $S_{\rm Wall}$}\label{sec:comparison}
As mentioned earlier, the constructions of $S_{\rm dyn}$ in \cite{Hollands:2024vbe} and $S_{\rm Wall}$ in \cite{Wall:2015raa, Bhattacharyya:2021jhr} are based on different arguments. Nevertheless, both constructions ultimately establish the second law for perturbations linearized around a stationary black hole solution. Both methods result in some modification of the definition of black hole entropy, involving terms that cannot be fixed by analyzing stationary situations alone. However, the modifications are not exactly equal; rather, they are related by a differential relation, as stated in \eqref{c0}.

$S_{\rm Wall}$ satisfies the second law provided that the black hole settles down to a stationary geometry in the asymptotic future, and the dynamical configuration remains sufficiently close to this final equilibrium, {\it i.e.}, the amplitude of the dynamics measures only a small deviation from the final stationary configuration. In \cite{Wall:2015raa, Bhattacharyya:2021jhr}, the authors have rearranged the higher derivative gravity equation to show that the rate of increase of $S_{\rm Wall}$ under any dynamics approaching equilibrium always decreases and ultimately vanishes at infinite future. Thus, this construction also justifies that $S_{\rm Wall}$ is the maximum of the entropy functional over the space of near-equilibrium dynamical geometries. 
  
On the other hand, to show the ever-increasing property of $S_{\rm dyn}$, one does not need to assume that the amplitude of the dynamics actually measures a small deviation from the final equilibrium. The construction of $S_{\rm dyn}$ is local in nature, in the sense that it increases immediately once an energy flux crosses the relevant null hypersurface (need not be the true event horizon). In contrast, due to the teleological nature of the event horizon, $S_{\rm Wall}$ changes in anticipation even before the dynamics have been turned on \cite{Visser:2024pwz}.

This qualitative distinction in the arguments leading to the second law is already well established, see \cite{Hollands:2024vbe}. In this section, however, we aim to examine the technical differences between the two approaches. The construction of $S_{\rm Wall}$ relies on a boost-weight analysis, which, strictly speaking, is not equivalent to the linearized amplitude expansion around a stationary background employed in the construction of $S_{\rm dyn}$. As a consequence, there exist certain terms that are neglected in one approach but retained in the other. In the following, we present an overview of these two techniques, with particular emphasis on the structural differences in their constructions.

\subsection{Arguments leading to the construction of $S_{\rm Wall}$}\label{subsec:Walle}
In this subsection, we briefly review the arguments used in \cite{Wall:2015raa, Bhattacharyya:2021jhr}. The key features of these arguments are as follows. First, it has been argued that without any loss of generality, around every null hypersurface, we could choose a  Gaussian null coordinate (GNC) system  such that the metric $g_{ab}$ takes the following form
\begin{equation}\label{metric1}
\begin{split}
ds^2 = 2\,dv\,dr - r^2 X\left(r,v,x^k\right)\,dv^2 + 2 r\,\omega_i\left(r,v,x^k\right)dv\,dx^i + h_{ij}\left(r,v,x^k\right)dx^i\,dx^j\,.
\end{split}
\end{equation}
Here, $r=0$ is the location of the null surface and $v$ is the affine parameter along its null generators. $x^i$ ($i=1,\cdots, D-2$) denotes the spatial coordinates along the constant $v$ and constant $r$ slices.
Next, it is noted that the form of the metric remains invariant under a residual gauge transformation
\begin{equation}
v\rightarrow v' = \lambda~v,~~~r\rightarrow r' = {r\over\lambda},~~~x^i\rightarrow x^{\prime i} =x^i\,.
\end{equation}
Under this coordinate transformation, any covariant tensor component transforms as
$$T^{b_1b_2\cdots b_k}_{a_1a_2\cdots a_m}\rightarrow {T'}^{b_1b_2\cdots b_k}_{a_1a_2\cdots a_m}= \lambda^w ~T^{b_1b_2\cdots b_k}_{a_1a_2\cdots a_m}$$
where $w$ is the boost weight of the tensor component, defined as
\begin{equation}
w =\text{(total number of lower $v$ and upper $r$ indices)} - \text{(total number of lower $r$ and upper $v$ indices)}\,.
\end{equation}
Also note that any covariant tensor constructed solely from the metric, when evaluated on the $r=0$ hypersurface, can depend only on the functions $X$, $\omega_i$, and $h_{ij}$, together with their $\partial_v$, $\partial_r$, and $\partial_i$ derivatives. Among these, only the derivatives $\partial_v$ and $\partial_r$ carry non-vanishing boost weights of $+1$ and $-1$ respectively.

From the above observation, it follows that if we evaluate any covariant tensor component of boost weight $w$, constructed solely out of metric, on $r=0$ hypersurface, we obtain
\begin{equation}
w = \text{total number of $\partial_v$ derivatives} - \text{total number of $\partial_r$ derivatives}\,.
\end{equation}
In other words, if we evaluate a tensor component of boost weight $w>0$, constructed solely out of metric, on the $r=0$ hypersurface, then in the final explicit expression, there will be exactly $w$ factors of $\partial_v$ derivatives that cannot be paired with any corresponding $\partial_r$ derivatives. We shall refer to each such excess $\partial_v$ derivative as an \textit{unpaired $\partial_v$ derivative}.

Now, for the purpose of entropy construction, we are interested in $E_{vv}$, the $vv$ component of the higher derivative equations of motion for the metric. This is clearly a covariant tensor component with boost weight $w=2$. So once evaluated on the $r=0$ surface, it could be decomposed as
\begin{equation}\label{evvst}
E_{vv}\vert_{r=0} = \sum_{k=1}^N {\mathfrak T}^{\{k\}}\,,
\end{equation}
where, ${\mathfrak T}^{\{k\}}$ denotes a term containing exactly $k$ positive boost-weight factors (see appendix \ref{app:notation} for the notational details). Each such term, therefore, takes the general form
\begin{equation}\label{tkst}
{\mathfrak T}^{\{k\}} = \sum_{\substack{m_1,\cdots,m_k\\ \sum_{k}m_k\geq2}}B_{\left(2-\sum_k m_k\right)}\prod_{i=1}^k\left[\partial_v^{m_i}A^{(i)}_{(0)}\right]\,,
\end{equation}
where all {\it unpaired $\partial_v$ derivatives} are written explicitly. The subscript index in parentheses denotes the boost weight, {\it e.g.},  $B_{\left(2-\sum_k m_k\right)}$ is a term with boost weight $\left(2-\sum_k m_k\right)$ and $A^{(i)}_{(0)}$ is a boost weight zero term for every $i$. Hence the product $\prod_{i=1}^k\left[\partial_v^{m_i}A^{(i)}_{(0)}\right]$ has boost weight $\sum_k m_k$ and each ${\mathfrak T}^{\{k\}}$ has boost weight $2$. 

It is very important to note that this structure of $E_{vv}$ as written in \eqref{evvst} is completely general and follows only from boost symmetry. It does not require any on-shell condition, {\it i.e.},  any explicit solution of the equations of motion. Furthermore, the off-shell decomposition of $E_{vv}$ in \eqref{evvst} is exact,  this decomposition is insensitive to whether or not one linearizes around a stationary black hole background\footnote{In the study of dynamical black hole entropy and the second law, one typically considers small perturbations about a stationary black hole background. The metric is then decomposed as
\begin{equation} \label{split_geometry}
    g_{ab} = \bar{g}_{ab}(rv, \vec x)+\epsilon \, \delta g_{ab}(r,v,\vec x) \,,
\end{equation}
where $\bar{g}_{ab}$ denotes the stationary background metric and $\epsilon$ parametrizes the amplitude of the perturbation. When $E_{vv}$ is evaluated on this decomposed geometry, the leading contribution to ${\mathfrak T}^{\{k\}}$ is of order ${\cal O}(\epsilon^k)$. More precisely,
\begin{equation}
{\mathfrak T}^{\{k\}}=\sum_{p=k}^{\infty}\epsilon^p\, {\cal T}_p(\bar g,\delta g)\,.
\end{equation}}.

In \cite{Bhattacharyya:2021jhr}, the authors analyzed the term ${\mathfrak T}^{\{1\}}$, {\it i.e.}, in the expansion of \eqref{evvst}, terms only involving a single positive boost weight factor, by exploiting the identity (see eq. (2.40) of \cite{Bhattacharyya:2021jhr})
\begin{equation}\label{ident}
2\,\xi^a\xi^bE_{ab}(g) + \xi_a\Theta^a(g,\mathcal{L}_\xi g) - \xi_a\xi^a L= \xi_a D_b Q^{ab}(g,\xi)\,,
\end{equation}
where, $D_a$ is covariant derivative with respect to the metric $g_{ab}$. Eqn. \eqref{ident} is a consequence of the diffeomorphism invariance of the gravitational Lagrangian density $L$, and therefore holds off-shell and for arbitrary vector fields $\xi$. Next, evaluating \eqref{ident} on the horizon $r=0$ with $\xi^a\partial_a = v\,\partial_v$, we obtain
\begin{equation}\label{identmid}
2\,v\sqrt{-g}\, E_{vv}(g) +\sqrt{-g}\,\Theta^r(g,\mathcal{L}_\xi g)-\partial_v\left[\sqrt{-g}\, Q^{rv}(g,\xi)\right] - \partial_i\left[\sqrt{-g}\, Q^{ri}(g,\xi)\right]=0\,.
\end{equation}
Analyzing different terms in \eqref{identmid} using their boost weight, one arrives at the following structure for ${\mathfrak T}^{\{1\}}$
\begin{equation}\label{t1}
{\mathfrak T}^{\{1\}} = -\partial_v\left[\frac{1}{\sqrt{h}}\partial_v\left(\sqrt{h}\,\mathcal{J}^v_{(0)}\right)+\nabla_i \mathcal{J}^{i}_{(1)}\right] +\tilde {\mathfrak T}^{\{2\}}
\end{equation}
where $\nabla_i$ denotes the covariant derivative with respect to $h_{ij}$. The term $\tilde{\mathfrak T}^{\{2\}}$ has the same structure as that of ${\mathfrak T}^{\{2\}}$, {\it i.e.}, it contains two unpaired $\partial_v$ derivatives acting on two different zero-boost terms. Finally, $\mathcal{J}^v_{(0)}$ has been identified with the Wall entropy $S_{\rm Wall}$ \cite{Wall:2015raa, Bhattacharyya:2021jhr, Hollands:2022ajj} as 
%\textcolor{blue}{NK: Please cross check the convention for the numerical factor in the following equation, maybe we cite some older works of ours or Wall.}
\begin{equation}\label{eq:Wallfnl}
S_{\rm Wall}=4\pi\int d^{D-2}x\,\sqrt{h}\,\mathcal{J}^v_{(0)}\,.
\end{equation}
We would again like to emphasize that the final form of the Wall entropy \eqref{eq:Wallfnl} is generated entirely through an off-shell rearrangement of the terms appearing in ${\mathfrak T}^{\{1\}}$, for which we have a clear prescription based on rearranging the unpaired $\partial_v$ operators in $E_{vv}$ evaluated on $r=0$ so as to cast it in the form \eqref{t1}. In particular,  the construction of $S_{\rm Wall}$ does not require any linearization around a stationary background, nor does it depend on any explicit $v$ coordinates as we have in the expression for $S_{\rm dyn}$ (see eqn. \eqref{defdyn}). However, as explained earlier, the fact that only the terms of the form ${\mathfrak T}^{\{1\}}$ in $E_{vv}$ have been analyzed can be physically motivated from the fact that these are precisely the only terms that contribute at linear order if the metric is decomposed into a stationary background and a small dynamical fluctuation, as in \cite{Hollands:2024vbe}.
 
In other words, in \cite{Bhattacharyya:2021jhr}, the authors have devised a technique to analyze terms with different boost weights by connecting them to the distribution of $\partial_v$ and $\partial_r$ operators. The authors have proven a particular structure for certain terms in $E_{vv}$ as a direct result of applying this technique. This structure turns out to be relevant in the physical situation of linearized dynamics around a stationary black hole geometry. But the linearization is not needed to prove the structure. This particular feature is the major technical difference from the method used to construct $S_{\rm dyn}$, as we shall see in the next subsection.

\subsection{Arguments leading to the construction of $S_{\rm dyn}$   }
In this subsection, we briefly review the arguments presented in \cite{Hollands:2024vbe} for the construction of $S_{\rm dyn}$, emphasizing the differences from the method used in \cite{Bhattacharyya:2021jhr}. The starting point of the analysis is the identity (see eqn.\,(2.39) of \cite{Bhattacharyya:2021jhr}) 
\begin{equation}\label{identdyn}
2\,\xi^bE^a_{b}(g) + \Theta^a(g,\mathcal{L}_\xi g) - \xi^a L= D_b Q^{ab}(g,\xi)\,.
\end{equation}
Note that eqn.\,\eqref{ident} is obtained by contracting \eqref{identdyn} with $\xi_a$.
In the next step, the metric is decomposed into a background part and a linear perturbation as $g_{ab} = \bar g_{ab} + \epsilon\,\delta g_{ab}$. Consequently, eqn.\,\eqref{identdyn} leads to
\begin{equation}\label{ident2}
\begin{split}
&2\,\xi^b\delta\left[\sqrt{-g}\,E^a_{b}(g)\right] +\delta\left[\sqrt{-g}\,\Theta^a(g,\mathcal{L}_\xi g)\right] - \xi^a\delta\left[\sqrt{-g}\, L\right]= \partial_b\Big(\delta\left[\sqrt{-g}\,Q^{ab}(g,\xi)\right]\Big)\,,
\end{split}
\end{equation}
where the variation of any object under the background plus fluctuation split \eqref{split_geometry} can be obtained as \begin{equation} \label{variation_def}
    \delta A[g]\equiv A[\bar g+\epsilon \,\delta g]  - A[\bar g] + {\cal O}(\epsilon^2) \, .
\end{equation} 
Next, assuming that $\bar g$ is stationary with $\xi$ as the corresponding Killing vector, one can manipulate the variation in \eqref{ident2} to obtain the following identity
\begin{equation}\label{ident3}
\begin{split}
&2\,\xi^b\delta\left[\sqrt{-g}\, E^a_{b}(g)\right] = \partial_b\bigg(\delta\left[\sqrt{-g}\, Q^{ab}(g,\xi) \right] + \sqrt{-\bar{g}}\,\xi^a\Theta^b(\bar g,\delta g) - \sqrt{-\bar{g}}\,\xi^b\Theta^a(\bar g,\delta g)\bigg)+\sqrt{-\bar{g}}\,\xi^a E^{bc}\delta g_{bc}\,.
\end{split}
\end{equation}
Note that the identity \eqref{ident3} holds only when $\bar g$ is stationary and terms of order ${\cal O}(\epsilon^2)$ are neglected.\\\\
Projecting the identity \eqref{ident3} along the Killing vector $\xi^a\partial_a = v\partial_v - r\partial_r$, and  restricting to the horizon, we obtain
\begin{equation}\label{iden4}
\begin{split}
2 \sqrt{-\bar g}\,v\,\delta E_{vv}\bigg\vert_{\rm Horizon}=\partial_v\bigg(\delta\left[\sqrt{-g}\, Q^{rv}(g,\xi) \right] - v\,\sqrt{-\bar g}\,\Theta^r(\bar g,\delta g)\bigg)+\partial_i \bigg(\delta\left[\sqrt{-g}\, Q^{ri}(g,\xi) \right]\bigg)\,.
\end{split}
\end{equation}
Both sides of \eqref{iden4} are of ${\cal O}(\epsilon)$ by construction. Integrating \eqref{iden4} over compact constant $v$ slices of the horizon (denoted as $\Sigma_v$) we find
\begin{equation}\label{iden5}
\begin{split}
2~\int_{\Sigma_v}d^{D-2}x\, \sqrt{-\bar g}\,v\,\delta E_{vv}\bigg\vert_{\rm Horizon}=\partial_v\bigg[\int_{\Sigma_v}d^{D-2}x\,\bigg(\delta\left[\sqrt{-g}\, Q^{rv}(g,\xi) \right] -v\,\sqrt{-\bar g}\,\Theta^r(\bar g,\delta g)\bigg)\bigg]\,.
\end{split}
\end{equation}
Now in \cite{Hollands:2024vbe}, the authors have shown that one could construct a geometric vector ${\cal B}^a$ so that
\begin{equation}\label{dynB}
\delta\left(\sqrt{-g}\,{\cal B}^r(g,\xi)\right)\bigg\vert_{\rm Horizon} =\sqrt{-\bar g}\,\Theta^r(\bar g,\delta g)\,.
\end{equation}
Substituting this relation in eqn.\,\eqref{iden5} one finds
\begin{equation}\label{iden6}
\begin{split}
2~\int_{\Sigma_v}d^{D-2}x\, \sqrt{-\bar g}~v~\delta E_{vv}\bigg\vert_{\rm Horizon}=\partial_v\int_{\Sigma_v}d^{D-2}x\,\delta\left[\sqrt{-g}~ \Big(Q^{rv}(g,\xi) - v\,{\cal B}^r(g,\xi)\Big) \right]\,.
\end{split}
\end{equation}
Finally $S_{\rm dyn}$ is identified as 
\begin{equation}\label{defdyn}
S_{\rm dyn}=2\pi\int_{\Sigma_v}d^{D-2}x\,\sqrt{h}\,\Big[Q^{rv}(g,\xi) - v\,{\cal B}^r(g,\xi)\Big]\,.
\end{equation}
Note that, unlike $S_{\rm Wald}$, the expression for $S_{\rm dyn}$, when evaluated in the coordinate system \eqref{metric1}, has explicit dependence on the coordinate $v$.

\subsection{A comparison between these two methods}\label{subsec:WallvsWald}
As explained earlier, when ${\mathfrak T}^{\{k\}}$ is evaluated on a split geometry $-$ decomposed into a stationary background $\bar g_{ab}$ and a dynamical fluctuation $\epsilon\,\delta g_{ab}$ $-$ the leading contribution appears at order ${\cal O}(\epsilon)^k$. Therefore, if the sum in eqn.\,\eqref{evvst} is truncated at $k=1$, the neglected terms are of order ${\cal O}(\epsilon)^2$ or higher. This seems to be the same approximation as used in \eqref{iden4}. 

However, recall that ${\mathfrak T}^{(1)}$ consists of terms containing exactly one positive boost-weight factor. These positive boost-weight factors are, in general, multiplied by appropriate negative or zero boost-weight factors (see \eqref{tkst} with $k=1$). Now, if a typical non-positive boost-weight factor is evaluated on the split geometry, it receives non-vanishing contributions at all orders in $\delta g$. The simplest example is the inverse metric, which carries zero boost weight. But we know that $(\bar g + \delta g)^{-1}$ is actually an infinite series in power of $\delta g$. In  the construction of $S_{\rm Wall}$, all such terms with non-positive boost weight are treated exactly \cite{Bhattacharyya:2021jhr}. 

On the other hand, for the construction of $S_{\rm dyn}$, the key equation that is used is \eqref{iden4}. This equation is correct only if $\delta$ denotes a deviation from a stationary metric $\bar g_{ab}$. Any term that could be expressed as Lie derivative of some other geometric quantity is considered to be of order ${\cal O}(\epsilon)$, even if such terms do not have any positive boost weight factor. For example, according to the analysis of \cite{Hollands:2024vbe}, a term of the form $(1 - v\partial_v) \partial_r h_{ij}= -\mathcal{L}_\xi \partial_r h_{ij}$ is treated as a term of order ${\cal O}(\epsilon)$, but this is a term of boost weight $-1$ and therefore treated exactly according to \cite{Bhattacharyya:2021jhr}.
 
Indeed, if we compare the identity \eqref{iden4} with identity \eqref{evvst}, evaluated on the split geometry, we see that the leading term - the order ${\cal O}(\epsilon)$ piece in $\left(2\sqrt{-\bar{g}}\,v\,{\mathfrak T}^{\{1\}}\right)$, matches with \eqref{iden4}. However in \cite{Bhattacharyya:2021jhr}, it has been  shown that the structure in identity \eqref{t1} is true for all terms in ${\mathfrak T}^{\{1\}}$ and not just the ${\cal O}(\epsilon)$ piece around equilibrium. In this sense, \eqref{iden4} is a special case of \eqref{t1}. On a split geometry, on the horizon, up to corrections of order ${\cal O}(\epsilon^2)$, we can write $2\sqrt{-\bar{g}}\,v\,{\mathfrak T}^{\{1\}}$ as follows
\begin{equation}\label{eq:ididt}
\begin{split}
2\sqrt{-\bar{g}}\,v\,{\mathfrak T}^{\{1\}}&=-2\,v\,\sqrt{h}\,\partial_v\left[\frac{1}{\sqrt{h}}\partial_v\left(\sqrt{h}\,\mathcal{J}^v_{(0)}\right)+\frac{1}{\sqrt{h}}\partial_i\left(\sqrt{h}\,\mathcal{J}^{i}_{(1)}\right)\right] +2\,v\,\sqrt{h}\,\tilde {\mathfrak T}^{\{2\}}\\
&=-2\,v\,\partial_v\left[\partial_v\left(\sqrt{h}\,\mathcal{J}^v_{(0)}\right)+\partial_i\left(\sqrt{h}\,\mathcal{J}^{i}_{(1)}\right)\right] +\mathcal{O}(\epsilon)^2\\
&=2\,\partial_v \bigg[\left(\sqrt{h}\,\mathcal{J}^v_{(0)}\right) - v \Big[\partial_v\left(\sqrt{h}\mathcal{J}^v_{(0)}\right) + \partial_i\left(\sqrt{h}\mathcal{J}^i_{(1)}\right)\Big]\bigg] +2\,\partial_i\left(\sqrt{h}\, \mathcal{J}^i_{(1)}\right)  +\mathcal{O}(\epsilon)^2\,.
\end{split}
\end{equation}
Since the left-hand sides of the eqs. \eqref{eq:ididt} and \eqref{iden4} are the same, we can equate the right-hand sides of the two equations, which leads to the following relation
\begin{equation}\label{rearrang}
\begin{split}
&2\,\partial_v \bigg[\left(\sqrt{h}\,\mathcal{J}^v_{(0)}\right) - v \Big[\partial_v\left(\sqrt{h}\mathcal{J}^v_{(0)}\right) + \partial_i\left(\sqrt{h}\mathcal{J}^i_{(1)}\right)\Big]\bigg] +2\,\partial_i\left(\sqrt{h}\, \mathcal{J}^i_{(1)}\right)\\
&=\partial_v\bigg(\delta\Big[\sqrt{h}\big[\, Q^{rv}(g,\xi)  - v\,\mathcal{B}^r( g,\xi)\big]\Big]\bigg)+\partial_i \bigg(\delta\left[\sqrt{h}\, Q^{ri}(g,\xi) \right]\bigg)
\end{split}
\end{equation}
where we have used the fact that, on the horizon, $\sqrt{-g}=\sqrt{h}$. 
Integrating both sides of \eqref{rearrang} on a constant $v$ slice of $r=0$ hypersurface and multiplying by $2\pi$ we find
\begin{equation}\label{rearrang2}
\begin{split}
&4\,\pi\,\partial_v\left[\int_{\Sigma_v}d^{D-2}x\,\sqrt{h}\, \mathcal{J}^v_{(0)} - v \partial_v \int_{\Sigma_v}d^{D-2}x\,\sqrt{h}\, \mathcal{J}^v_{(0)}\right]\\
&=2\,\pi\,\partial_v\bigg(\delta\int_{\Sigma_v}d^{D-2}x\,\sqrt{h}\,\Big[ Q^{rv}(g,\xi) - v~{\cal B}^r(g,\xi)\Big] \bigg) +{\cal O}(\epsilon^2)\,.
\end{split}
\end{equation}
Using the definitions of $S_{\rm Wall}$ and $S_{\rm dyn}$ given in eqs. \eqref{eq:Wallfnl} and \eqref{defdyn}, respectively, we obtain
\begin{equation}
\begin{split}
&\partial_v\left[(1-v\partial_v)S_{\rm Wall}\right] = \partial_v \delta S_{\rm dyn}+{\cal O}(\epsilon^2)\\
\Rightarrow~&\partial_v\left[S_{\rm dyn}-(1-v\partial_v)S_{\rm Wall}\right] ={\cal O}(\epsilon^2)\,.
\end{split}
\end{equation}
In the last line we have used the fact that $\partial_v S_{\rm dyn}$ vanishes on any stationary solution and therefore $\partial_v S_{\rm dyn} =\partial_v \delta S_{\rm dyn} + {\cal O}(\epsilon^2) $.

Now we could use the fact that on a stationary solution both $S_{\rm Wall}$ and $S_{\rm dyn}$ must agree, leading to the relation \eqref{c0}
\begin{equation}
S_{\rm dyn}-(1-v\partial_v)S_{\rm Wall} ={\cal O}(\epsilon^2)\,.
\end{equation}
If we use the language of \cite{Bhattacharyya:2021jhr}, then both $S_{\rm Wall}$ and $S_{\rm dyn}$ have total boost weight zero.  By construction both of them can have at most one factor with positive boost weight, multiplied by other factors with balancing negative boost weight. We already know that $S_{\rm Wall}$ does not have any explicit $v$ dependence whereas $S_{\rm dyn}$ does; so in case of $S_{\rm dyn}$, the negative boost weight terms also include explicit powers of $v$.

Apart from terms with one positive boost weight factor, both $S_{\rm dyn}$ and $S_{\rm Wall}$ also contain terms with  zero positive boost weight factors. These are the terms that do not have any unpaired $\partial_v$ or $\partial_r$ derivatives (or explicit powers of $v$ in case of $S_{\rm dyn}$). In other words,
$S_{\rm dyn}$ and $S_{\rm Wall}$ could be decomposed as
\begin{equation}\label{decomp}
\begin{split}
&S_{\rm dyn} = S_{\rm dyn}^{\rm zero} + S_{\rm dyn}^{\rm rest}, \quad \text{and,}\quad S_{\rm Wall} = S_{\rm Wall}^{\rm zero} + S_{\rm Wall}^{\rm rest}
\end{split}
\end{equation}
where $S_{\rm dyn}^{\rm zero}$ and $S_{\rm wall}^{\rm zero}$ consist of those terms that have zero positive boost weight factors. These are the term that will be non-zero in stationary black holes. It follows that
  \begin{equation}\label{decompeq1}
\begin{split}
S_{\rm dyn}^{\rm zero} = S_{\rm Wall}^{\rm zero}\,.
\end{split}
\end{equation}
Hence the relation \eqref{c0} reduces to 
\begin{equation}\label{decompeq2}
\begin{split}
&S_{\rm dyn}^{\rm rest}= -v\,\partial_v\, S_{\rm Wall}^{\rm zero} +(1 - v\,\partial_v) S_{\rm Wall}^{\rm rest} + {\cal O}(\epsilon^2)\,.
\end{split}
\end{equation}
Clearly, any error term (the ${\cal O}(\epsilon^2)$ terms in \eqref{decompeq2}) must consist of a single positive boost-weight factor multiplied by suitable negative boost-weight factor(s) (or explicit powers of $v$) such that the total boost weight vanishes. However, upon evaluating these terms on the split geometry, they must be of order ${\cal O}(\epsilon^2)$ or higher. This implies that, in every error term, at least one of the negative boost-weight factors must be acted upon by the operator
\begin{equation}
\mathcal{L}_\xi\equiv (-n + v\,\partial_v),\qquad n>0
\end{equation}
where $(-n)$ denotes the boost weight of the corresponding factor.

\section{Constructing $S_{\rm Wall}$ from $S_{\rm dyn}$}\label{sec:Swall}
The derivation of the relation \eqref{c0} between $S_{\rm Wall}$ and $S_{\rm dyn}$ works provided we assume the existence of local expressions (ones that do not involve any $v$ integration) for both $S_{\rm Wall}$ and $S_{\rm dyn}$ (see \S \ref{subsec:WallvsWald}). To elaborate,
\begin{itemize}
\item Wald's covariant phase space argument has been used to arrange a particular component of the equations of motion to get an expression for $S_{\rm dyn}$ or more precisely the first order dynamical fluctuation of it. 
\item Then using the boost weight argument (or some analogue of it as done in section VI of \cite{Hollands:2024vbe}), one could show that the same component could be rearranged to generate the first order fluctuation of $S_{\rm Wall}$ . 
\item Since it is the same component of the equations of motion that is being rearranged, and since the first-order fluctuation of the equations of motion around any stationary geometry is unique, the relation \eqref{c0} follows.
\end{itemize}
In this section we would like to see how we could directly derive $S_{\rm Wall}$ starting with an expression of $S_{\rm dyn}$ by solving the relation \eqref{c0} around some arbitrary stationary solution. Note that, after the integration over the spatial coordinates, both $S_{\rm dyn}$ and $S_{\rm Wall}$ are functions of $v$ only (since both of them are evaluated at $r=0$). Therefore, given $S_{\rm dyn}$, the equation \eqref{c0} could be viewed as an inhomogeneous ODE for $S_{\rm Wall}$. We could solve it  by simple integration as follows
\begin{equation}\label{sols}
\begin{split}
S_{\rm Wall}(v) &= -v\int dv\left(S_{\rm dyn}(v)\over{ v}^2\right) + C\,v +{\cal O}(\epsilon^2)\\
&= S_{\rm dyn}(v) + v\int dv\left(\partial_v S_{\rm dyn}(v)\over v\right) + C\,v+{\cal O}(\epsilon^2) \, ,
\end{split}
\end{equation}
where $C$ is an arbitrary integration constant. 

 This solution for $S_{\rm Wall}$ will be analytic around $v=0$ ({\it i.e}, all order $v$ derivatives are finite at $v=0$), provided $\partial_v S_{\rm dyn}\vert_{v=0}$ vanishes, since otherwise $S_{\rm Wall}$ will be non-analytic because of a term proportional to $\log(v)$. The vanishing of the first $v$ derivative of $S_{\rm dyn}$ at $v=0$ simply follows from equation \eqref{iden6} by substituting $v=0$ and assuming that $\delta E_{vv}\vert_{v=0}$ is finite.
 
However, in this section, we would like to see whether, given an $S_{\rm dyn}$, one can derive a local expression for $S_{\rm Wall}$ that does not involve any integration over $v$. This is possible provided \eqref{c0} is identically true for every dynamics up to linear order in the amplitude of the fluctuation or every possible form of $\delta g$. In other words, we would like to solve the relation \eqref{c0} algebraically using a combination of boost weight analysis and amplitude expansion for both $S_{\rm Wall}$ and $S_{\rm dyn}$, such that it works for arbitrary off-shell $v$ dependence of $\delta g$.
 
\subsection{Algorithm to algebraically fix $S_{\rm Wall}$ from $S_{\rm dyn}$}\label{subsec:algorithm}
First, let us introduce some notation: any tensor component with boost weight $w>0$, and only one positive boost weight factor, could be expressed as 
\begin{equation}
T_{(w)} = \sum_{m=0}^M A_{(-m)}B_{(m+w)}\,.
\end{equation}
As mentioned before, in the construction of $S_{\rm Wall}$, all negative and zero boost weight factors are treated exactly. If $\epsilon$ denotes the amplitude of dynamical fluctuation, then the negative boost weight factor $A_{(-m)}$ does contain terms of all orders in $\epsilon$.
The linearization in $\epsilon$ could be ensured by imposing the condition that all factors with negative boost weight are evaluated on a stationary geometry and therefore,  on the horizon, satisfy
equations of the form
\begin{equation}
(v\partial_v -m)A_{(-m)}|_{r=0}= 0\quad\Rightarrow \quad \partial_v\left[v^{-m}A_{(-m)}\right]_{r=0} =0\quad\,\,\Rightarrow \quad A_{(-m)}|_{r=0} = v^m\,A^{[m]} \, ,
\end{equation}
where, $A^{[m]}$ is independent of $v$. So, a positive boost weight tensor component, which is also linear in the amplitude of fluctuation, will admit the following expansion in powers of $v$
\begin{equation}\label{eq:exph}
T_{(w)} = \sum_{m=0}^M A_{(-m)}B_{(m+w)} = \sum_{m=0}^Mv^m A^{[m]}B_{(m+w)}\,,
\end{equation}
where $A^{[m]} \equiv A^{[m]}(rv =0)$ denotes a term,  evaluated on the horizon of the equilibrium geometry (and therefore a function of the product $rv$) and with at least $m$ derivatives with respect to the product $rv$. According to this notation, all $A^{[m]}$ are independent of $v$, whereas $B_{(m)}$ generally carries an implicit $v$-dependence arising from the dynamical perturbation $\delta g$, even when $m=0$ (see Appendix \ref{app:notation} for further details on the notation). We expect the expansion \eqref{eq:exph} to be closely related to the expansion used in eqn. (53) of the paper \cite{Hollands:2024vbe}.\\\\
Using this notation both $S_{\rm Wall}$ and $S_{\rm dyn}$ admit the following expansions
\begin{equation}\label{eq:decomp}
\begin{split}
&S_{\rm Wall} \equiv \sum_{m=0}\,S_{(-m)}T_{(m)}=\sum_{m=0}v^m\,S^{[m]}T_{(m)} +{\cal O}(\epsilon)^2 \, ,\\
&S_{\rm dyn} \equiv\sum_{m=0}v^m\,\tilde S^{[m]}\tilde T_{(m)}+{\cal O}(\epsilon)^2 \, .
\end{split}
\end{equation}
Note that, unlike $S_{\rm Wall}$, the expression $S_{\rm dyn}$ contains explicit powers of $v$ from the outset (see eqn. \eqref{defdyn}). Consequently, there is little advantage in expressing it as a product of negative and positive boost weight tensor components. With the above decomposition \eqref{eq:decomp}, the relation \eqref{c0} takes the form
\begin{equation}\label{eq:c0exp}
\begin{split}
0&=(1- v\partial_v) S_{\rm Wall}-S_{\rm dyn}\\
&=S^{[0]}T_{(0)} +\sum_{m=1}v^m\left[(1-m)S^{[m]}T_{(m)} - S^{[m-1]}\partial_v T_{(m-1)}\right] -\sum_{m=0}v^m\,\tilde S^{[m]}\tilde T_{(m)}\,.
\end{split}
\end{equation}
We expect eqn.\,\eqref{eq:c0exp} to be satisfied identically for every dynamics and for this to be true  the coefficients  of $v^m$ for every $m$ must vanish individually. At this stage we should emphasize again that the coefficients themselves are functions of $v$, since the $v$ dependence of the positive boost weight factors are completely determined by the metric fluctuation - $\delta g$ which we have kept arbitrary. Equating the coefficients of different powers of $v$, we obtain
\begin{equation}\label{eq:vpower}
\begin{split}
v^0:~~&\tilde S^{[0]}\tilde T_{(0)} = S^{[0]}T_{(0)} \, ,\\
v^1: ~~ &\tilde S^{[1]}\tilde T_{(1)}  = -S^{[0]}\partial_v T_{(0)}= -\partial_v\left(S^{[0]}T_{(0)}\right)= -\partial_v\left(\tilde S^{[0]}\tilde T_{(0)} \right) \, .
\end{split}
\end{equation}
Note that the second equation in \eqref{eq:vpower} is actually an equation for the coefficients of $S_{\rm dyn}$ itself. In fact, it simply says that for every dynamical fluctuation of small amplitude
\begin{equation}\label{newrel}
\partial_v S_{\rm dyn}\vert_{v=0} = \tilde S^{[1]}\tilde T_{(1)} +\partial_v\left(\tilde S^{[0]}\tilde T_{(0)} \right)= 0\,.
\end{equation}
This is the same condition that follows from the defining equation of $S_{\rm dyn}$, namely that $v\,\delta E_{vv}\sim \partial_v S_{\rm dyn}$ (see eqn. \eqref{iden6}).

Next, we would like to show that if the condition in the second equation of \eqref{eq:vpower} is satisfied, we could determine a local $S_{\rm Wall}$ without any further new constraints. Equating the coefficients of a general $v^n\,(n\geq 2)$, we find
\begin{equation}\label{vgen}
\begin{split}
v^n:~~&\tilde S^{[n]}\tilde T_{(n)} =(1-n) S^{[n]}T_{(n)}-\partial_v\left[S^{[n-1]}T_{(n-1)}\right],~~~n\geq2 \, .
\end{split}
\end{equation}
Suppose we are dealing with a $2N+2$ derivative theory. Then $S_{\rm Wall}$ will have $2N$ derivatives. Therefore, the maximum value of $n$ could be $N$. As a result $(1-v\partial_v)S_{\rm Wall}$ could have maximum power $v^{N+1}$ and the coefficient will have the structure $-\partial_v\left[S^{[N]}T_{N}\right]$. So finally, the set of equations \eqref{vgen} and \eqref{eq:vpower} become
%In other words, the equation we get by equating the coefficients of $v^{N+1}$ will have the form
%\begin{equation}\label{vlast}
%\begin{split}
%v^{N+1}:~~&\tilde S^{(N+1)}\tilde T_{(N+1)}   =-\partial_v\left[S^{(N)}T_{(N)}\right]
%\end{split}
%\end{equation}
%So finally the set of equations
\begin{equation}\label{vall}
\begin{split}
v^{N+1}:~~&\tilde S^{[N+1]}\tilde T_{(N+1)}   =-\partial_v\left[S^{[N]}T_{(N)}\right]~~~N\geq 1 \, ,\\
v^{N}:~~&\tilde S^{[N]}\tilde T_{(N)}  =(1-N) S^{[N]}T_{(N)}-\partial_v\left[S^{[N-1]}T_{(N-1)}\right] \, ,\\
&\vdots\\
v^{2}:~~&\tilde S^{[2]}\tilde T_{(2)}  =- S^{[2]}T_{(2)}-\partial_v\left[S^{[1]}T_{(1)}\right] \, ,\\
v^{1}:~~ &\tilde S^{[1]}\tilde T_{(1)}  = -\partial_v\left(S^{[0]}T_{(0)}\right)= -\partial_v\left(\tilde S^{[0]}\tilde T_{(0)}\right) \, ,\\
v^0:~~&\tilde S^{[0]}\tilde T_{(0)}  = S^{[0]}T_{(0)} \,.\\
\end{split}
\end{equation}
Now we start solving from the top equation in \eqref{vall}. By comparing the coefficient of $v^{N+1}$ we solve for $S^{[N]}T_{(N)}$. We could find a local solution for $S^{[N]}T_{(N)}$ provided we rewrite the LHS as a total $\partial_v$ derivative. But this is always possible since according to our notation, $\tilde T_{(N+1)}$ has $(N+1)$ unpaired $\partial_v$ derivatives and $\tilde S^{[N+1]}$ is $v$-independent. Once we found $S^{[N]}T_{(N)}$, we substitute this solution into the next equation and solve for $S^{[N-1]}T_{(N-1)}$. We can continue this procedure recursively to determine all $S^{[m]}T_{(m)}$ up to $m=1$, {\it i.e.}, up to the coefficient of $v^2$. At this stage, we have two equations, corresponding to the coefficients of $v^0$ and $v^1$, for the single unknown $\tilde S^{[0]}\tilde T_{(0)}$. One of these equations can therefore be used to determine $\tilde S^{[0]}\tilde T_{(0)}$, while the other must be satisfied as a consistency condition.

In summary, if the condition we found by comparing the coefficients of $v^0$ and $v$ is satisfied, there is no further obstruction in constructing a local $S_{\rm Wall}$ from $S_{\rm dyn}$ up to linear order in the amplitude of fluctuation.

\subsection{Comparing \eqref{newrel} with relation (3.10) of \cite{Bhattacharyya:2021jhr}}
In \cite{Bhattacharyya:2021jhr}, the authors have proved the existence of $S_{\rm Wall}$ using the boost weight analysis of a particular component of the equations of motion. A crucial intermediate result in that proof is the relation between the Noether charge $Q$ and the presymplectic potential $\Theta$, given in eqn. (3.10) of that work. In this section, we demonstrate that the relation \eqref{newrel} arises as a special case of the more general identity derived there.

For convenience, let us first quote the eqn. (3.10) from \cite{Bhattacharyya:2021jhr} along with a translation to notations used in the present note and an integration over the compact spatial slice $\Sigma_v$ of the horizon at an arbitrary constant $v$.
\begin{equation}\label{eq:2105}
\begin{split}
&\Theta_{1} - W^{rv}_v - \partial_v \tilde Q_{rv} = \int_{\Sigma_v}\sqrt{h}\,d^{D-2}x\,\nabla_i \mathcal{J}^i_{(1)}=0\,.
\end{split}
\end{equation}
where different terms are defined as follows (see eqn. (3.7) in \cite{Bhattacharyya:2021jhr}). Here we have implicitly assumed that $\Theta^r$ and $Q^{rv}$ are integrated over $\Sigma_v$
\begin{equation}\label{2105not}
\begin{split}
\int_{\Sigma_v}\sqrt{h}\,d^{D-2}x\,\Theta^r(g,\mathcal{L}_\xi g)\vert_{r=0} &\equiv \Theta_{1} + v \Theta_{2},\qquad\int_{\Sigma_v}\sqrt{h}\,d^{D-2}x\,Q^{rv}(g,\xi)\vert_{r=0} \equiv \tilde Q^{rv} + v~W^{rv}_v,\\
&Q^{ri}(g,\xi)\vert_{r=0} \equiv \mathcal{J}^i_{(1)} + v\partial_v \tilde{\mathcal{J}}^i_{(1)}\,.
\end{split}
\end{equation}
Using the notation introduced earlier for the expansion about equilibrium, we express $\Theta_1$, $W^{rv}_{v}$, and $\tilde{Q}^{rv}$ in the following form:
\begin{equation}\label{eq:qtheta}
\begin{split}
&\tilde Q^{rv} \equiv \sum_{m=0}v^m\,P^{[m]}Q_{(m)},\qquad W^{rv}_v \equiv \sum_{m=0}v^m \,V^{[m]}W_{(m+1)}\\
&~~~~~~~~~~~~~~~~~~~~~~~\Theta_1 \equiv\sum_{m=0}v^m\alpha^{[m]}\beta_{(m+1)}\,.
\end{split}
\end{equation}
Now substituting the expansion \eqref{eq:qtheta} to the relation \eqref{eq:2105} we find
\begin{equation}\label{eq:2105next}
\begin{split}
\sum_{m=0}v^m\left[\alpha^{[m]}\beta_{(m+1)} - V^{[m]} W_{(m+1)} - (m+1)P^{[m+1]} Q_{(m+1)} - \partial_v\left(P^{[m]}Q_{(m)}\right)\right]=0\,.
\end{split}
\end{equation}
Now, using the arguments of \cite{Bhattacharyya:2021jhr}, it follows that eqn.\,\eqref{eq:2105next} holds identically. Consequently, the coefficient of $v^m$ must vanish identically for every value of $m$. In particular, setting the coefficient of $v^0$ to zero yields
\begin{equation}\label{eq:rel1}
\begin{split}
\alpha^{[0]}\beta_{(1)} = V^{[0]} W_{(1)} +P^{[1]} Q_{(1)} + \partial_v\left(P^{[0]}Q_{(0)}\right)\,.
\end{split}
\end{equation}
$S_{\rm dyn}$ can also be expressed in terms of $Q^{rv}$ and another geometric quantity defined on the horizon, namely ${\cal B}^r$ (see eqn.\,\eqref{defdyn}). Since ${\cal B}^r$ carries boost weight $(+1)$, it admits an expansion around equilibrium analogous to that given in \eqref{eq:qtheta}
\begin{equation}
\int_{\Sigma_v}\sqrt{h}\,d^{D-2}x\,{\cal B}^r = \sum_{m=0}v^m b^{[m]}c_{(m+1)}\,.
\end{equation}
Accordingly, we write
\begin{equation}\label{expB}
\begin{split}
S_{\rm dyn}&=2\pi\int_{\Sigma_v}d^{D-2}x\,\sqrt{h}\,\Big[Q^{rv}(g,\xi) - v\,{\cal B}^r(g,\xi)\Big]\\
&=2\pi\,P^{[0]}Q_{(0)}+2\pi\sum_{m=1}v^m\left[P^{[m]}Q_{(m)} +V^{[m-1]}W_{(m)}-b^{[m-1]}c_{(m)}\right]\,.
\end{split}
\end{equation}
$S_{\rm dyn}$ admits another expansion \eqref{eq:decomp}
\begin{equation}\label{eq:decomp22}
\begin{split}
&S_{\rm dyn} \equiv\sum_{m=0}v^m\,\tilde S^{[m]}\tilde T_{(m)}\,.
\end{split}
\end{equation}
Comparing \eqref{expB} with \eqref{eq:decomp22}, we obtain
\begin{equation}
\begin{split}
\tilde S^{[0]}\tilde T_{(0)} &= 2\pi\,P^{[0]}Q_{(0)}\\
\tilde S^{[m]}\tilde T_{(m)} &= 2\pi\Big(P^{[m]}Q_{(m)} +V^{[m-1]}W_{(m)}-b^{[m-1]}c_{(m)}\Big)~~~m\geq1\,.
\end{split}
\end{equation}
The relation \eqref{newrel} reduces to
\begin{equation}\label{newrel2}
P^{[1]}Q_{(1)} +V^{[0]}W_{(1)}-b^{[0]}c_{(1)}+\partial_v\left(P^{[0]}Q_{(0)}\right) =0\,.
\end{equation}
The above relation \eqref{newrel2} will be compatible with the
relation in  \eqref{eq:rel1}, provided
\begin{equation}\label{tobe}
\alpha^{(0)}\beta_{(1)} = b^{(0)}c_{(1)}\,.
\end{equation}
Eqn.\,\eqref{tobe} could be argued as follows.
We know that ${\cal B}^r(g,\xi)$ is defined so that around equilibrium
$$\delta {\cal B}^r(g,\xi) \equiv  {\cal B}^r(\bar g+ \delta g,\xi)-{\cal B}^r(\bar g,\xi) + {\cal O}(\delta g)^2= \Theta^r(\bar g,\delta g)\,.$$
Now, the $\Theta^r$ that appears in equation (3.10) of \cite{Bhattacharyya:2021jhr} is defined as $\Theta^r(g,\delta g = \mathcal{L}_\xi g)$. To arrive at eqn. \eqref{eq:rel1} we have to first expand  $\Theta^r(g, \delta g =\mathcal{L}_\xi g)$ around the equilibrium
\begin{equation}\label{iden7}
\begin{split}
\delta \Theta^r(g, \mathcal{L}_\xi g) = \Theta^r(\bar g,  \mathcal{L}_\xi \delta g) =\mathcal{L}_\xi \left[\Theta^r(\bar g,  \delta g)\right] =(1 + v\partial_v) \Theta^r(\bar g,  \delta g)=(1 + v\partial_v) \delta {\cal B}^r(g,\xi)\,.
\end{split}
\end{equation}
Again \eqref{iden7} should be true identically for any dynamical fluctuation $\delta g$. Hence, if we substitute the expansion \eqref{expB} and \eqref{eq:qtheta} in \eqref{iden7}, coefficients of each power of $v$ should match. The matching of the coefficient of $v^0$ gives the relation \eqref{tobe}.

\section{An explicit example: Riemann-squared theory}\label{sec:example}
In this section, we explicitly illustrate our construction in the context of Riemann-squared theory. We first show how the entropy current discussed in \cite{Bhattacharyya:2021jhr} arises from the construction of \cite{Hollands:2024vbe}, particularly from eqn.\,\eqref{iden4}. We then verify the relations \eqref{decompeq1} and \eqref{decompeq2}. Finally, we demonstrate that the dynamical entropy $S_{\rm dyn}$ satisfies the consistency condition \eqref{newrel}. Along the way, we also verify the relation \eqref{c0}, which was previously verified in \cite{Visser:2024pwz} for a general $f(\rm Riemann)$ theory.
\subsection{Entropy current from \cite{Hollands:2024vbe}}
We start from eqn. \eqref{iden4}
\begin{equation}\label{eq:evvex}
\begin{split}
2 \sqrt{-\bar g}\,v\,\delta E_{vv}\bigg\vert_{\rm Horizon}=\partial_v\bigg(\delta\left[\sqrt{-g}~ \Big(Q^{rv}(g,\xi) - v\,{\cal B}^r(g,\xi)\Big) \right]\bigg)+\partial_i \bigg(\delta\left[\sqrt{-g}\, Q^{ri}(g,\xi) \right]\bigg)\,,
\end{split}
\end{equation}
where we have used the relation \eqref{dynB}. Here, we quote the final expressions for $Q^{rv}, Q^{ri}$ and $\mathcal{B}^r$; the details of their derivation are presented in Appendix \ref{app:Riemann}
\begin{equation}\label{eq:QQB}
\begin{split}
Q^{rv}&=-8\,v\Big[(\partial_v+K)R^{rvrv}-\bar{K}_{ij}R^{rjri}-K_{ij}R^{jvri}+\nabla_i R^{rvri}\Big]+8 R^{rvrv}+\mathcal{O}(\epsilon^2)\\
Q^{ri}&=8R^{rirv}-8v\,\partial_v R^{rirv}-8v\,\nabla_j R^{rirj}-8v\,\omega_j\,R^{rirj}+\mathcal{O}(\epsilon^2)
\end{split}
\end{equation}
\begin{equation}\label{eq:Brdef}
\sqrt{-\bar{g}}\,\Theta^r(\bar{g},\delta g)=\delta\Big[\sqrt{-g}\,8\,R^{rjiv}K_{ij}\Big]\,.
\end{equation}
The last eqn. \eqref{eq:Brdef} implies
\begin{equation}
\mathcal{B}^r=8\,R^{rjiv}K_{ij}\,.
\end{equation}
Substituting these expressions in \eqref{eq:evvex}, we obtain
\begin{equation}\label{eq:impeqn}
\begin{split}
&2 \sqrt{-\bar g}\,v\,\delta E_{vv}\bigg\vert_{\rm Horizon}\\
&=\partial_v\bigg(\delta\left[\sqrt{-g}~ \Big(-8v\left[(\partial_v+K)R^{rvrv}-\bar{K}_{ij}R^{rjri}-K_{ij}R^{jvri}\right]+8 R^{rvrv} - 8v\,R^{rjiv}K_{ij}\Big) \right]\bigg)\\
&-\partial_v\bigg(\delta\Big[\sqrt{-g}\,8v\nabla_i R^{rvri}\Big]\bigg)+\partial_i \bigg(\delta\left[ \sqrt{-g}\,\Big(8R^{rirv}-8v\,\partial_v R^{rirv}-8v\,\nabla_j R^{rirj}-8v\,\omega_j\,R^{rirj}\Big) \right]\bigg)\,.
\end{split}
\end{equation}
The last line of the above equation simplifies to
\begin{equation}
\begin{split}
&-\partial_v\bigg(\delta\Big[\sqrt{-g}\,8v\nabla_i R^{rvri}\Big]\bigg)+\partial_i \bigg(\delta\left[ \sqrt{-g}\,\Big(8R^{rirv}-8v\,\partial_v R^{rirv}-8v\,\nabla_j R^{rirj}-8v\,\omega_j\,R^{rirj}\Big) \right]\bigg)\\
&\equiv 2v\sqrt{-g}\,\partial_v\left[\frac{1}{\sqrt{-g}}\partial_i\left(\sqrt{-g}\,\delta \mathcal{J}^i\right)\right]
\end{split}
\end{equation}
where,
\begin{equation}
\mathcal{J}^i=-4\,h^{ij}\partial_v \omega_j+4\,\nabla_j K^{ij}\,.
\end{equation}
This expression precisely matches with the entropy current obtained for Riemann-squared theory in \cite{Bhattacharyya:2021jhr} (see Table 7 therein). Note that, $\mathcal{J}^i$ gets contribution both from $Q^{rv}$ and $Q^{ri}$. We now multiply \eqref{eq:impeqn} by $2\pi$ and integrate it over a compact constant-$v$ cross-section of the horizon. Since the $\mathcal{J}^i$ term is a total divergence, its integral vanishes, and we obtain
\begin{equation}\label{eq:comprc}
\begin{split}
&4\pi\int_{\Sigma_v} d^{D-2}x \sqrt{-\bar g}\,v\,\delta E_{vv}\bigg\vert_{\rm Horizon}\\
&=2\pi\int_{\Sigma_v} d^{D-2}x\, \partial_v\bigg(\delta\left[\sqrt{-g}~ \Big(-8v\left[(\partial_v+K)R^{rvrv}-\bar{K}_{ij}R^{rjri}-K_{ij}R^{jvri}\right]+8 R^{rvrv} - 8v\,R^{rjiv}K_{ij}\Big) \right]\bigg)\,.
\end{split}
\end{equation}
Comparing \eqref{eq:comprc} with \eqref{iden6} and \eqref{defdyn} we find
\begin{equation}\label{eq:Rimdyn}
\begin{split}
S_{\rm dyn}&=2\pi \int d^{D-2} x\sqrt{h}\bigg[-8\,v\left[(\partial_v+K)R^{rvrv}-\bar{K}_{ij}R^{rjri}-K_{ij}R^{jvri}\right]+8 R^{rvrv}-8\,v\,R^{rjiv}K_{ij}\bigg]\\
&=16\pi\int d^{D-2} x\sqrt{h}\bigg[-v(\partial_v+K)R^{rvrv}+v\bar{K}_{ij}R^{rjri}+R^{rvrv}\bigg]\\
&=16\pi\int d^{D-2} x\bigg[(1-v\partial_v)\left(\sqrt{h}R^{rvrv}\right)+v\sqrt{h}\,\bar{K}_{ij}R^{rjri}\bigg]\,,
\end{split}
\end{equation}
where we have used $\sqrt{-g}\,|_{r=0}=\sqrt{h}$. Now, using the expression of $R^{rjri}$ from appendix C of \cite{Bhattacharya:2019qal}, we obtain
\begin{equation}\label{eq:fdyn}
\begin{split}
S_{\rm dyn}&=16\pi(1-v\,\partial_v)\int d^{D-2} x\, \sqrt{h}\Big[R^{rvrv}+\bar{K}^{ij} K_{ij}\Big]+16\pi\int d^{D-2} x\,\sqrt{h} K_{ij}\left(\bar{K}^{ij}-v\partial_v\bar{K}^{ij}\right)\\
&=16\pi(1-v\,\partial_v)\int d^{D-2} x\, \sqrt{h}\Big[R^{rvrv}+\bar{K}^{ij} K_{ij}\Big]\,,
\end{split}
\end{equation}
where, in the last line we have used the identity $v\partial_v \bar{K}^{ij}\Big|_{r=0}=\bar{K}^{ij}\Big|_{r=0}+{\cal O}(\delta g)$.\\
On the other hand, Wall entropy as defined in \eqref{eq:Wallfnl} takes the following form
\begin{equation}\label{eq:rimwall}
S_{\rm Wall}=16\pi\int d^{D-2} x\, \sqrt{h}\Big[R^{rvrv}+\bar{K}^{ij} K_{ij}\Big]\,,
\end{equation}
where we have used the expression of $\mathcal{J}^v$ as given in Table 7 of \cite{Bhattacharyya:2021jhr}
\begin{equation}
\mathcal{J}^v=4\,R_{rvrv}+4\bar{K}^{ij}K_{ij}\,.
\end{equation}
Note that the expression for $\mathcal{J}^v$ carries an overall minus sign relative to the corresponding result in \cite{Bhattacharyya:2021jhr}. This difference arises because eqn.\,\eqref{t1} contains an additional minus sign compared to the convention adopted in \cite{Bhattacharyya:2021jhr}. Comparing \eqref{eq:rimwall} with \eqref{eq:Rimdyn}, we obtain
\begin{equation}
S_{\rm dyn}=(1-v\partial_v)S_{\rm Wall}\,,
\end{equation}
which is the relation \eqref{c0}.
\subsection{Verifying relations \eqref{decompeq1} and \eqref{decompeq2}}
$S_{\rm dyn}$ is decomposed as eqn. \eqref{decomp}
\begin{equation}
S_{\rm dyn} = S_{\rm dyn}^{\rm zero} + S_{\rm dyn}^{\rm rest}\,.
\end{equation}
Using \eqref{eq:fdyn}, we find
\begin{equation}\label{eq:dync}
\begin{split}
S_{\rm dyn}^{\rm zero}&=16\pi\int d^{D-2} x\, \sqrt{h}\,R^{rvrv}\\
S_{\rm dyn}^{\rm rest}&=-16\pi\, v\partial_v\int d^{D-2} x\, \sqrt{h}\,R^{rvrv}+16\pi(1-v\,\partial_v)\int d^{D-2} x\, \sqrt{h}\bar{K}^{ij} K_{ij}\,.
\end{split}
\end{equation}
To make this more explicit, we note that for the Riemann-squared theory the relevant curvature component takes the form $R^{rvrv}=X+\omega^2/4$. Similarly, $S_{\rm Wall}$ is also decomposed as eqn. \eqref{decomp}
\begin{equation}
S_{\rm Wall} = S_{\rm Wall}^{\rm zero} + S_{\rm Wall}^{\rm rest}\,.
\end{equation}
Using eqn. \eqref{eq:rimwall}, we find
\begin{equation}\label{eq:wallc}
\begin{split}
S_{\rm Wall}^{\rm zero}&=16\pi\int d^{D-2} x\, \sqrt{h}R^{rvrv}\,,\\
S_{\rm Wall}^{\rm rest}&=16\pi\int d^{D-2} x\, \sqrt{h}\bar{K}^{ij} K_{ij}\,.
\end{split}
\end{equation}
Comparing \eqref{eq:dync} and \eqref{eq:wallc}, we obtain
\begin{equation}
\begin{split}
S_{\rm dyn}^{\rm zero}&= S_{\rm Wall}^{\rm zero}\,,\\
S_{\rm dyn}^{\rm rest}&= -v\,\partial_v\, S_{\rm Wall}^{\rm zero} +(1 - v\,\partial_v) S_{\rm Wall}^{\rm rest} + {\cal O}(\epsilon^2)\,.
\end{split}
\end{equation}
The above are precisely the relations, \eqref{decompeq1} and \eqref{decompeq2}, that we wanted to verify.
\subsection{Constructing $S_{\rm Wall}$ from $S_{\rm dyn}$}
$S_{\rm dyn}$ has the expression \eqref{eq:fdyn}
\begin{equation}\label{eq:fss}
\begin{split}
S_{\rm dyn}&=16\pi(1-v\,\partial_v)\int d^{D-2} x\, \sqrt{h}\Big[R^{rvrv}+\bar{K}^{ij} K_{ij}\Big]\,.
\end{split}
\end{equation}
From this expression, we want to construct the corresponding $S_{\rm Wall}$ following the algorithm discussed in subsection \S \ref{subsec:algorithm}. First, we simplify \eqref{eq:fss} as follows
\begin{equation}\label{eq:cmpp1}
\begin{split}
S_{\rm dyn}&=16\pi(1-v\,\partial_v)\int d^{D-2} x\, \sqrt{h}\Big[\Big(X+\frac{\omega^2}{4}\Big)-\frac{1}{4}\left(\partial_r h^{ij}\right)(\partial_v h_{ij})\Big]\\
&=16\pi\int d^{D-2}x\bigg[\sqrt{h}\Big(X+\frac{\omega^2}{4}\Big)-v\partial_v\left(\sqrt{h}\Big[X+\frac{\omega^2}{4}\Big]\right)-\frac{1}{4}\sqrt{h}\left(\partial_r h^{ij}\right)(\partial_v h_{ij})\\
&~~~~~~~~~~~~~~~~~~~~~+\frac{1}{4}\sqrt{h}v\left(\partial_v\partial_r h^{ij}\right)(\partial_v h_{ij})+\frac{1}{4}\sqrt{h}v\left(\partial_r h^{ij}\right)(\partial_v^2 h_{ij})\bigg]\\
&=16\pi\int d^{D-2}x\bigg[\sqrt{h}\Big(X+\frac{\omega^2}{4}\Big)-v\partial_v\left(\sqrt{h}\Big[X+\frac{\omega^2}{4}\Big]\right)-\frac{1}{4}\sqrt{h}v\left( h^{\prime ij}\right)(\partial_v h_{ij})\\
&~~~~~~~~~~~~~~~~~~~~~+\frac{1}{4}\sqrt{h}v\left(h^{\prime ij}\right)(\partial_v h_{ij})+\frac{1}{4}\sqrt{h}v^2\left(h^{\prime ij}\right)(\partial_v^2 h_{ij})\bigg]\,,
\end{split}
\end{equation}
where, $`\,\prime$\,' denotes derivative with respect to the product $r v$. Now, we expand $S_{\rm dyn}$ as eqn. \eqref{eq:decomp}
\begin{equation}\label{eq:cmpp2}
\begin{split}
&S_{\rm dyn} \equiv\sum_{m=0}v^m\,\tilde S^{[m]}\tilde T_{(m)}+{\cal O}(\epsilon)^2\,.
\end{split}
\end{equation}
Comparing \eqref{eq:cmpp1} with \eqref{eq:cmpp2}, we obtain
\begin{equation}\label{eq:ref100}
\begin{split}
\tilde S^{[0]}\tilde T_{(0)}&=16\pi\int d^{D-2}x\,\sqrt{h}\Big(X+\frac{\omega^2}{4}\Big)\\
\tilde S^{[1]}\tilde T_{(1)}&=16\pi\int d^{D-2}x\bigg[-\partial_v\left(\sqrt{h}\Big[X+\frac{\omega^2}{4}\Big]\right)\bigg]\\
\tilde S^{[2]}\tilde T_{(2)}&=16\pi\int d^{D-2}x\bigg[\frac{1}{4}\sqrt{h}\left(h^{\prime ij}\right)(\partial_v^2 h_{ij})\bigg]\,.
\end{split}
\end{equation}
From, eqn. \eqref{eq:ref100} we see that
\begin{equation}\label{eq:conver}
\tilde S^{[1]}\tilde T_{(1)} +\partial_v\left(\tilde S^{[0]}\tilde T_{(0)} \right)= 0\,,
\end{equation}
which is the consistency condition \eqref{newrel}. Now consider eqn. \eqref{vall}. Since we are working with the Riemann-squared theory, which is a four-derivative theory, we have $N=1$. We can write $\tilde S^{[2]}\tilde T_{(2)}$ as follows
\begin{equation}
\begin{split}
\tilde S^{[2]}\tilde T_{(2)}&=16\pi\int d^{D-2}x\bigg[\frac{1}{4}\sqrt{h}\left(h^{\prime ij}\right)(\partial_v^2 h_{ij})\bigg]\\
&=16\pi\int d^{D-2}x\,\partial_v\bigg[\frac{1}{4}\sqrt{h}\left(h^{\prime ij}\right)(\partial_v h_{ij})\bigg]\,.
\end{split}
\end{equation} 
From the first equation in \eqref{vall}, which in the present case corresponds to the coefficient of $v^2$, we obtain
\begin{equation}\label{eq:wallp1}
\begin{split}
S^{[1]} T_{(1)}&=-16\pi\int d^{D-2}x\,\bigg[\frac{1}{4}\sqrt{h}\left(h^{\prime ij}\right)(\partial_v h_{ij})\bigg]\,.
\end{split}
\end{equation} 
The coefficient of $v$ corresponds to the consistency condition \eqref{newrel} which we have verified in \eqref{eq:conver}. From the $v^0$ coefficient, we obtain
\begin{equation}\label{eq:wallp2}
S^{[0]} T_{(0)}=16\pi\int d^{D-2}x\,\sqrt{h}\Big(X+\frac{\omega^2}{4}\Big)\,.
\end{equation}
Adding \eqref{eq:wallp1} and \eqref{eq:wallp2}, we obtain
\begin{equation}
\begin{split}
S_{\rm Wall}&=16\pi\int d^{D-2}x\,\sqrt{h}\bigg[\Big(X+\frac{\omega^2}{4}\Big)-\frac{1}{4}v\left(h^{\prime ij}\right)(\partial_v h_{ij})\bigg]\\
&=16\pi\int d^{D-2}x\,\sqrt{h}\bigg[R^{rvrv}+\bar{K}^{ij} K_{ij}\bigg]\,.
\end{split}
\end{equation}
This precisely matches the expression for the Wall entropy given in \eqref{eq:rimwall}.

\section{Discussions}\label{sec:disc}
In this note, we have made a comparative study of two approaches to constructing black hole entropy in higher-curvature theories of gravity, namely, the Wall entropy developed in \cite{Wall:2015raa, Bhattacharyya:2021jhr} and the dynamical entropy developed in \cite{Hollands:2024vbe}. We have established that these two constructions differ in the underlying approximation schemes.  In \cite{Hollands:2024vbe}, a perturbative expansion (up to linear order) in the small fluctuation around a stationary solution is used, whereas \cite{Bhattacharyya:2021jhr} uses boost weight analysis, where all terms with negative or zero boost weights have been treated exactly. As a consequence, the Wall entropy generally contains contributions that are higher order in the dynamical fluctuations, even though it is not expected to provide the correct notion of black hole entropy beyond linear order (see Appendix \ref{App:Toy} for a simple toy example).

One interesting feature of these two constructions is that the defining equation for $S_{\rm dyn}$ involves a single $\partial_v$ derivative \eqref{eq:sdynfd}, whereas the defining equation for $S_{\rm Wall}$ involves two $\partial_v$ derivatives \eqref{eq:swlf}. Consequently, in the case of the Wall entropy, the null energy condition implies not only that the entropy increases, but also that the rate of entropy increase decreases monotonically until the system reaches the final equilibrium state, where this rate vanishes. Thus, the equilibrium state corresponds to the state of maximum entropy.
 
It is known that entropy production occurs at the quadratic order in amplitude.  Up to the linear order, all we can establish is that the entropy is not destroyed, or in other words, entropy does not change. Neither method can make a definitive prediction about the sign of the net entropy production beyond linear order. So the key question at this stage is how to extend these constructions systematically to the next orders. Several recent works have investigated black hole entropy beyond linearized perturbation theory \cite{Ashtekar:2026jdz,Fu:2026itf}.

The boost-weight analysis provides a relatively straightforward framework for such extensions within the effective field theory approach, as demonstrated in \cite{Davies:2023qaa}. Its main drawback, however, is that it is closely tied to a particular choice of coordinates. As a result, the resulting entropy functional may fail to possess the desired covariance properties, as discussed in \cite{Hollands:2024vbe}. In contrast, covariance is built into the construction of the dynamical entropy. However, extending this framework systematically beyond linear order appears to be considerably more challenging.

A promising direction for future research is therefore to combine the strengths of these two approaches: the systematic higher-order organization afforded by the boost-weight analysis and the manifest covariance of the dynamical entropy construction. Such a synthesis may ultimately provide a route toward a non-perturbative understanding of black hole thermodynamics in higher-derivative theories of gravity.

The teleological nature of the event horizon implies that the Wall entropy is intrinsically nonlocal, whereas the dynamical entropy is a local quantity. In \cite{Hollands:2024vbe}, it was shown that, in Einstein gravity, the dynamical entropy is naturally identified with the area of the apparent horizon, this was subsequently extended to $f(R)$ theories in \cite{Jia:2025tgf}. An important direction for future work is to determine whether a similar correspondence persists in higher-curvature theories of gravity. Moreover, there has been substantial progress in understanding the laws of black hole thermodynamics for quasilocal horizons \cite{Ashtekar:2004cn,Ashtekar:2026jdz}. Understanding the connection (if any) between dynamical entropy and quasilocal horizons in higher-curvature gravity is therefore an interesting and important avenue for future investigation.

\section*{Acknowledgements} 
We would like to thank Prateksh Dhivakar and Shuvayu Roy for useful discussions during the initial stages of this project. We acknowledge the use of ChatGPT solely for language editing and improving the clarity of selected portions of this manuscript. SB and NK would like to acknowledge the warm hospitality from TIFR, Mumbai, and The Lodha Mathematical Sciences Institute (LMSI) during an academic visit while this work was in progress. SB would also like to acknowledge the warm hospitality of SINP and IISER Pune during the course of this work. NK acknowledges support from a recently concluded MATRICS research grant (MTR/2022/000794) from the Anusandhan National Research Foundation (ANRF), India. Finally, we express our sincere gratitude to the people of India for their steady support of basic scientific research.
 
\appendix

\section{Notations}\label{app:notation}
\begin{itemize}
\item A \emph{subscript index} in parentheses, $A_{(n)}$, denotes the boost weight of the corresponding term; for example, $A_{(n)}$ has boost weight $n$.
%\item A lower-case index in parentheses, $A_{(n)}$, denotes the boost weight of the corresponding term; {\it e.g.}, $A_{(n)}$ has boost weight $n$. 

\item A \emph{superscript index} in curly brackets, $A^{\{n\}}$, denotes the number of positive boost-weight factors in a term; {\it e.g.}, $A^{\{n\}}$ contains {\it exactly} $n$ positive boost-weight factors.

\item A \emph{superscript index} in square brackets, $A^{[n]}$, denotes the \textit{minimum} number of derivatives with respect to the product $rv$ evaluated at the horizon $r = 0$, where $A$ represents a term constructed from the equilibrium metric. Therefore $A^{[n]}\equiv A^{[n]}(rv=0,x^i)$ has {\it atleast} $n$ derivatives with respect to the product $rv$. 
To clarify this notation further, $A^{[n]}$ is defined through the relation
\begin{equation}\label{eq:sqno}
A_{(-n)} \equiv v^n A^{[n]}_{(0)} .
\end{equation}

The quantity $A_{(-n)}$ must contain $n$ uncontracted $\partial_r$ derivatives in excess of the contracted $\partial_r \partial_v$ derivatives. Each uncontracted derivative acting on the equilibrium metric produces a factor of $v$, which explains the overall factor $v^n$.

On the other hand, the contracted $\partial_r \partial_v$ derivatives, when evaluated on the horizon, do not generate additional factors of $v$. However, they still introduce extra derivatives with respect to the product $rv$. For this reason, the index $[n]$ on the right-hand side of \eqref{eq:sqno} should strictly be interpreted as the \textit{minimum} number of derivatives rather than the exact number.

A more precise notation would therefore be $A^{[\ge n]}$. Nevertheless, for the purposes of the present discussion, this distinction is not important. To avoid unnecessary clutter, we will continue to use the simpler notation $A^{[n]}$.
 
\item $D_a$ denotes the covariant derivative with respect to the horizon-adapted metric $g_{ab}$, {\it i.e.}, the metric given in \eqref{metric1}. On the other hand, $\nabla_i$ denotes the covariant derivative with respect to the induced metric $h_{ij}$ on constant-$v$ slices of the horizon. 

\item {\it Definitions:}
\begin{equation}
\begin{split}
&K_{ij}=\frac{1}{2}\partial_v h_{ij},\qquad \bar{K}_{ij}=\frac{1}{2}\partial_r h_{ij},\qquad K^{ij}=-\frac{1}{2}\partial_v h^{ij},\qquad \bar{K}^{ij}=-\frac{1}{2}\partial_r h^{ij},\\
&K=h^{ij}K_{ij}=\frac{1}{\sqrt{h}}\partial_v\sqrt{h},\qquad \bar{K}=h^{ij}\bar{K}_{ij}=\frac{1}{\sqrt{h}}\partial_r\sqrt{h}\,.
\end{split}
\end{equation}

%An upper-case index in square brackets, $A^{[n]}$, denotes the {\it minimum} number of derivatives with respect to the product $rv$ at the horizon $r=0$, where $A$ is a term constructed from the equilibrium metric. Let us explain this notation a bit further, $A^{[n]}$ is defined through the relation
%\begin{equation}\label{eq:sqno}
%A_{(-n)}\equiv v^n A^{[n]}_{(0)}
%\end{equation}
%Now, $A_{(-n)}$ has to have $n$ uncontracted $\partial_r$ derivatives over contracted $\partial_r\partial_v$ derivatives, each of these uncontracted derivatives acting on the equilibrium metric brings a factor of $v$ giving rise to the factor of $v^n$. But the contracted $\partial_r\partial_v$ derivatives when evaluated on the horizon though does not produce any extra factor of $v$ still give rise to extra derivatives with respect to the product $rv$ that is the reason in the RHS of \eqref{eq:sqno} $[n]$ denotes the minimum number of derivatives. Therefore a better notation would be $A^{[\geq n]}$, but for our discussion the fact that $[n]$ denotes the minimum number of derivatives is not important. Therefore we will stick with the notation $A^{[n]} \equiv A^{[n]}(rv=0,x^i)$ to avoid clumsyness.
\end{itemize}

\section{Verifying \eqref{iden7} in four-derivative theories of gravity}
In this appendix, we will verify eqn. \eqref{iden7}
\begin{equation}
\begin{split}
\delta \Theta^r(g, \mathcal{L}_\xi g) =(1 + v\partial_v) \delta {\cal B}^r(g,\xi) 
\end{split}
\end{equation}
for four derivative theories of gravity.

Let us begin with the Riemann-squared theory $(L=R^{abcd}R_{abcd})$. We can get the expression of ${\cal B}^r$ from \cite{Hollands:2024vbe} where the authors have written explicitly the dualized version of ${\cal B}^a$ which is a $n-1$ form $\mathbf{B}_{\cal H}$ for four derivative theories of gravity (see eqn. (98) in \cite{Hollands:2024vbe}). For Riemann squared theory, the $(n-1)$ form $\mathbf{B}_{\cal H}$ is
\begin{equation}
(B_{\cal H})_{a_1\cdots a_{n-1}}=N^f \epsilon_{f a_1\cdots a_{n-1}}4\, R^{abcd}\xi_a N_d \mathcal{L}_\xi g_{bc}\,,
\end{equation}
where $N^a$ is the auxiliary null vector field that satisfies $N^a\xi_a=1$ on the horizon. On the horizon, $\xi^a \partial_a=v\partial_v$ which implies $N_a dx^a=(1/v)dv$. We can also write $\xi_a dx^a=v\, dr$ and $N^a\partial_a=(1/v)\partial_r$. Therefore, we have
\begin{equation}
\begin{split}
(B_{\cal H})_{a_1\cdots a_{n-1}}=\frac{1}{v}\, \epsilon_{r a_1\cdots a_{n-1}}4\, R^{rbcv} \mathcal{L}_\xi g_{bc}\,.
\end{split}
\end{equation}
Using \eqref{eq:Drxi23} in the expression
\begin{equation}
\mathcal{L}_\xi g_{bc}=D_b\xi_c+D_c\xi_b,
\end{equation}
we find that the only non-vanishing component of the Lie derivative is $\mathcal{L}_\xi g_{ij}=2v\,K_{ij}$ which gives
\begin{equation}
(B_{\cal H})_{a_1\cdots a_{n-1}}=8\,\epsilon_{r a_1\cdots a_{n-1}}\, R_{vijr}K^{ij}\,.
\end{equation}
Therefore the dualized vector is given by
\begin{equation}
{\cal B}^r = 8 R_{vijr}K^{ij}\,,
\end{equation}
which gives
\begin{equation}\label{eq:compp1}
\begin{split}
\delta {\cal B}^r(g,\xi) &\equiv  {\cal B}^r(\bar g+ \delta g,\xi)-{\cal B}^r(\bar g,\xi) + {\cal O}(\delta g)^2\\
&=8 R_{vijr}\,\delta K^{ij}\,.
\end{split}
\end{equation}
On the other hand, from Table~6 of \cite{Bhattacharyya:2021jhr}, we have the following expression for the Riemann-squared theory:
\begin{equation}\label{th1}
\begin{split}
\Theta^r(g,\mathcal{L}_\xi g) &= 8(1+ v\partial_v)\Big[R_{vijr}K^{ij} 
+K_{ij}\partial_v\bar K^{ij}-\partial_v\left(\bar K_{ij} K^{ij}\right) \Big]+ 8v~\partial_v^2(\bar K^{ij}K_{ij})\\
&=8(1+ v\partial_v)\left[R_{vijr}K^{ij}\right]-8\,\bar{K}_{ij}\partial_v K^{ij}+8v(\partial_v \bar{K}_{ij})(\partial_v K^{ij})\,,
\end{split}
\end{equation}
where, we have neglected terms with more than one positive boost weight factors.
Now, 
\begin{equation}\label{eq:compp2}
\begin{split}
\delta \Theta^r(g, \mathcal{L}_\xi g)&= \Theta^r(\bar g,  \mathcal{L}_\xi \delta g)\\
&=8(1+ v\partial_v)\left[R_{vijr}\delta K^{ij}\right]-8\,\bar{K}_{ij}\partial_v \delta K^{ij}+8v(\partial_v \bar{K}_{ij})(\partial_v \delta K^{ij})+\mathcal{O}(\delta g)^2\\
&=8(1+ v\partial_v)\left[R_{vijr}\delta K^{ij}\right]\,.
\end{split}
\end{equation}
In the last line, we have used the fact that, on the stationary background geometry,
\begin{equation}
v\partial_v \bar{K}^{ij}\Big|_{r=0}
=
\bar{K}^{ij}\Big|_{r=0}
+
{\cal O}(\delta g).
\end{equation}
Comparing \eqref{eq:compp1} with \eqref{eq:compp2}, we find
\begin{equation}
\delta \Theta^r(g, \mathcal{L}_\xi g) =(1 + v\partial_v) \delta {\cal B}^r(g,\xi) \,,
\end{equation}
which is the desired relation \eqref{iden7}.\\\\
Now, we will move on to the Ricci-tensor-squared theory. After some simplifications, we obtain the following expression for the Ricci-tensor-squared theory:
\begin{equation}
\begin{split}
(B_{\cal H})_{a_1\cdots a_{n-1}}&=N^f \epsilon_{f a_1\cdots a_{n-1}}\big(-R^{bc}-R_{ad}\xi^a N^d g^{bc}\big)\mathcal{L}_\xi g_{bc}\\
&=-2\,\epsilon_{r a_1\cdots a_{n-1}}\big(R^{ij}K_{ij}+K\, R_{vr}\big)\,.
\end{split}
\end{equation}
Therefore the dualized vector is given by
\begin{equation}\label{eq:RiT1}
\begin{split}
{\cal B}^r &= -2\big(R^{ij}K_{ij}+K\, R_{vr}\big)\\
\Rightarrow\,\, \delta {\cal B}^r&=-2\big(R^{ij}\delta K_{ij}+R_{vr}\,\delta K\big)\,.
\end{split}
\end{equation}
On the other hand, from Table~6 of \cite{Bhattacharyya:2021jhr}, we have the following expression for the Ricci tensor-squared theory:
\begin{equation}\label{eq:RiT2}
\begin{split}
\Theta^r(g,\mathcal{L}_\xi g) &= 2(1+ v\partial_v)\left[-R^{ij}K_{ij} -R_{rv}K +K\partial_v\bar K -\partial_v(K\bar K)\right]+ 2v~\partial_v^2(K\bar K)\\
&=2(1+ v\partial_v)\left[-R^{ij}K_{ij} -R_{rv}K\right]+2v(\partial_v K)(\partial_v\bar{K})-2(\partial_v K)\bar{K}\\
&=2(1+ v\partial_v)\left[-R^{ij}K_{ij} -R_{rv}K\right]\\
\Rightarrow\,\,\delta\Theta^r(g,\mathcal{L}_\xi g)&=2(1+ v\partial_v)\left[-R^{ij}\delta K_{ij} -R_{rv}\delta K\right]\,.
\end{split}
\end{equation}
Comparing \eqref{eq:RiT1} with \eqref{eq:RiT2}, we find
the relation \eqref{iden7}.\\\\
Similarly, for Ricci scalar squared theory we have
\begin{equation}\label{eq:RiS1}
\begin{split}
(B_{\cal H})_{a_1\cdots a_{n-1}}&=-2\,N^f \epsilon_{f a_1\cdots a_{n-1}}R\,g^{bc}\mathcal{L}_\xi g_{bc}\\
&=-4\,\epsilon_{r a_1\cdots a_{n-1}}R\,K\,.
\end{split}
\end{equation}
On the other hand, from Table~6 of \cite{Bhattacharyya:2021jhr}, for Ricci scalar squared theory, we have
\begin{equation}\label{eq:RiS2}
\Theta^r=-4(1+v\partial_v)(R\, K)\,.
\end{equation}
Comparing \eqref{eq:RiS1} with \eqref{eq:RiS2}, we find
\begin{equation}
\delta \Theta^r(g, \mathcal{L}_\xi g) =(1 + v\partial_v) \delta {\cal B}^r(g,\xi)\,,
\end{equation}
which is the desired relation \eqref{iden7}.

\section{Clarifying the physical process version of the first law for the zero-boost terms in $E_{vv}$}\label{App:PPF}

In this appendix, we clarify the physical-process version of the first law by focusing explicitly on how the differences between the constructions of $S_{\rm dyn}$ and $S_{\rm Wall}$ affect its validity and application. 

Previously, from the boost weight analysis, we learned that $E_{vv}\vert_{r=0}$ must have the off-shell structure of \eqref{evvst}. Building on this, without using any identity of the form \eqref{ident}, a simple rearrangement of $\partial_v$ allows us to show that
\begin{equation} \label{App_C_eq_1}
    {\mathfrak T}^{(1)} = \underbrace{\partial_v\left(A_{(0)}\partial_v B_{(0)}\right)}_\text{zero boost term} + \partial_v^2\left(\sum_{m=1}A_{(-m)}B_{(m)}\right) \, , \qquad {\mathfrak T}^{(1)}_\text{zero-boost} \equiv \partial_v\left(A_{(0)}\partial_v B_{(0)}\right) \, .
\end{equation}
The first term on the RHS of the above identity has been called the zero-boost-weight term, ${\mathfrak T}^{(1)}_\text{zero-boost}$. The key question is whether we can pull out another overall $\partial_v$ from ${\mathfrak T}^{(1)}_\text{zero-boost}$. Achieving this would allow us to write the RHS of the identity in \eqref{App_C_eq_1} as $\partial^2_v$ acting on some object. This step is clearly possible if we split the metric into a stationary background and dynamical fluctuations, and keep only terms up to order $\mathcal{O}(\epsilon)$. In such cases, within the zero-boost term, $A_{(0)}$ must be evaluated on the stationary metric, enabling $\partial_v$ to pass through $A_{(0)}$ and yielding ${\mathfrak T}^{(1)}_\text{zero-boost} \sim \partial^2_v \left(A_{(0)}B_{(0)}\right)$. 

However, if we do not substitute such a split at this stage, both $A_{(0)}$ and $B_{(0)}$ can have non-trivial $v$ dependence via the generic dynamical metric components. In such a general setting, ${\mathfrak T}^{(1)}_\text{zero-boost}$ cannot be written as an overall $\partial^2_v$ acting on some object. Importantly, none of the terms in ${\mathfrak T}^{(1)}$, including $A_{(0)}$ and $B_{(0)}$, have any explicit dependence on any of the coordinates. With this understanding in place, we can now turn to the implications for the physical-process version of the first law discussed in \cite{Wall:2015raa}. 

According to \cite{Wall:2015raa}, if one is not able to pull out another overall $\partial_v$ from the zero boost terms, {\it i.e.}, ${\mathfrak T}^{(1)}_\text{zero-boost}$, then this would lead to contradictions with the physical process version of the first law. To clarify how this issue arises, let us proceed by assuming the contrary: that both $\partial_v$ derivatives can not be pulled out as an overall $\partial_v^2$ for the zero-boost term ${\mathfrak T}^{(1)}_\text{zero-boost}$. 

In the physical process version of the first law, we need to integrate $v\, E_{vv}$ (where $E_{vv}$ denotes the $vv$-component of the field equations) on the horizon between two constant $v$ slices, $v = v_i$ and $v = v_f$, where the choice of slices satisfies $(v_i \ll v_0 - \delta v < v_0 + \delta v \ll v_f)$. Here, the infalling matter source, which represents the nonzero energy-momentum flux, is non-zero only within the interval $(v_0 - \delta v)$ to $(v_0 + \delta v)$. To state the physical process version of the first law, we manipulate the integration of $v\, E_{vv}$ on the horizon between these two constant $v$-slices, after integrating over the other spatial coordinates:

\begin{equation}\label{ppfl1}
\begin{split}
\int_{v_i}^{v_f} dv~v~E_{vv}\big\vert_{r=0;~\text{zero-boost} }
\sim& \int_{v_i}^{v_f} dv~v~\partial_v\left(A_{(0)}\partial_v B_{(0)}\right)\\
\sim&\int_{v_i}^{v_f} dv~\left[\partial_v~\left(v~A_{(0)}\partial_v B_{(0)}\right)-\left(A_{(0)}\partial_v B_{(0)}\right)\right]\\
\sim&\left[\left(v~A_{(0)}\partial_v B_{(0)}\right)\right]_{v_i}^{v_f}-\int_{v_i}^{v_f} dv~\left[\left(A_{(0)}\partial_v B_{(0)}\right)\right]\\
\sim&-\int_{v_i}^{v_f} dv~\left[\left(A_{(0)}\partial_v B_{(0)}\right)\right] \, ,
\end{split}
\end{equation}
The first term in the second-to-last line vanishes because it must be evaluated either long before the dynamical perturbation was turned on or long after it has vanished. In both scenarios, the black hole is assumed to settle to a stationary state where any positive boost term (terms containing $\partial_v$ not paired with $\partial_r$) will vanish.

Next, by splitting the metric as a sum of a stationary part plus a dynamical fluctuation and ignoring all terms non-linear in the amplitude of the dynamics, we evaluate $A_{(0)}$ on the $v$-independent stationary metric. Consequently, $A_{(0)}$ comes out of the integration, leading to
\begin{equation}
    \int_{v_i}^{v_f} dv~ \left(A_{(0)}\partial_v B_{(0)}\right) = A_{(0)}^{\rm eq}\int_{v_i}^{v_f}\partial_v B_{(0)} =A_{(0)}^{\rm eq} \left[B_{(0)}\right]_{v_i}^{v_f} \, .
\end{equation}
With this at hand, we can now analyze the implications for the description of entropy changes. It should be noted that in this discussion, two equilibrium black hole geometries are involved $-$ the initial and the final ones. In the above expression, $A_{(0)}^{eq}$ must be evaluated around one of these equilibrium configurations, and then multiplied by the difference of $B_{(0)}$ at these two equilibrium geometries. Thus, the final expression cannot be interpreted as a total change in a single geometric quantity, which could be identified with an expression for entropy differences between the initial and final equilibrium states. In particular, expanding around the initial or final equilibrium gives two different answers, simply because $A_{(0)}\vert_\text{initial BH}\neq A_{(0)}\vert_\text{final BH}$.

On the other hand, if we could write the zero-boost term of $E_{vv}$ as an overall $\partial_v^2$ acting on an object, say, $E_{vv} \sim -\partial_v^2 S_{(0)}$, then the steps from \eqref{ppfl1} would lead to:
$$\int_{v_i}^{v_f} dv~v~E_{vv}\sim \left[S_{(0)}\right]_{v_i}^{v_f} \sim\delta(\text{entropy}) \, ,$$
where we could then identify $S_{(0)}$ as the entropy. 

Now, in Wald's formalism, we have the following relation (all spatial integrations are suppressed)
$$v~ \delta E_{vv}\sim \partial_v\delta S_{\rm dyn}(v) \, $$
which can lead us to 
$$\int_{v_i}^{v_f}dv~v~ \delta E_{vv}\sim \delta S_{\rm dyn}(v_f)-\delta S_{\rm dyn}(v_i) \, ,$$
where for any geometric quantity $A[g]$, the notation $\delta A$ is defined as 
$$\delta A[\bar g,\delta g] \equiv A[\bar g + \delta g] - A[\bar g] + {\cal O}(\delta g^2)\, ,$$
such that $\bar g$ denotes a stationary geometry.

Therefore, the question of which equilibrium (initial or final) to expand around also arises in Wald's formalism, since it deals with only those terms involving the variation $\delta$. However, $S_{\rm dyn}$ can be defined in any geometry. Thus, the final result will be independent of which equilibrium is chosen for expanding around. To explain this in more detail, suppose we expand around the initial equilibrium:
\begin{equation}\label{ppfl2}
\begin{split}
&\delta S_{\rm dyn}(v_f) -\delta S_{\rm dyn}(v_i) =  S_{\rm dyn}[\bar g_{\rm initial} + \delta g(v_f)] - S_{\rm dyn}[\bar g_{\rm initial} + \delta g(v_i)] + {\cal O}(\delta g^2) \, ,
\end{split}
\end{equation}
where
\begin{equation}
\begin{split}
&\delta S_{\rm dyn}\equiv S_{\rm dyn}[\bar g_{\rm initial} + \delta g] - S_{\rm dyn}[\bar g_{\rm initial}]+ {\cal O}(\delta g)^2  \, .
\end{split}
\end{equation}
By our assumption, $\delta g(v_i) =0$ because this is the metric perturbation before matter passes through the horizon. Also, $\bar g_{\rm initial} + \delta g(v_f)=\bar g_{\rm final}$ because the black hole reaches equilibrium by that time. Thus,\footnote{There is an implicit assumption that $\bar g_{\rm final} -\bar g_{\rm initial} = {\cal O}(\delta g)$.}
\begin{equation}
\delta S_{\rm dyn}(v_f) -\delta S_{\rm dyn}(v_i) =  S_{\rm dyn}[\bar g_{\rm final} ]- S_{\rm dyn}[\bar g_{\rm initial}] + {\cal O}(\delta g^2) \, .
\end{equation}
Expanding around the final equilibrium leads to the same answer. In this case, we substitute $\delta g(v_f) =0$ and $\delta g(v_i)\neq 0$.

Therefore, to summarize, Wald's formalism tells us that the difference in entropy between two `nearby' stationary black holes connected by dynamics of small amplitude could be expressed as the change in some geometric quantity evaluated at the two stationary end points. It follows that it is irrelevant which equilibrium we choose to expand the dynamics around. Consequently, it would be in direct contradiction with Wall's prescription, had we not been able to write the zero-boost terms as $\partial_v^2 S_{(0)}$ with an overall $\partial_v^2$ outside.

\section{Calculational details for Section \S \ref{sec:example}}\label{app:Riemann}
In this appendix, we provide some details of the computations of $\Theta^r$, $Q^{rv}$ and $Q^{ri}$
\subsection*{Computation of $\Theta^r$}
For the Riemann-squared theory, the symplectic potential $\Theta^\mu$ is given by the following expression (see Table 4 of \cite{Bhattacharyya:2021jhr}):
\begin{equation}\label{eq:expThm}
\Theta^\mu=4 R^{\mu\rho\nu\sigma}D_\sigma \delta g_{\nu\rho}-4(D_\sigma R^{\mu\nu\rho\sigma})\delta g_{\nu\rho}\,.
\end{equation}
We are working in the horizon adapted coordinate system \eqref{metric1}
\begin{equation}
ds^2 = 2\,dv\,dr - r^2X\left(r,v,x^k\right)dv^2 + 2\,r\,w_i\left(r,v,x^k\right)dv\,dx^i + h_{ij}\left(r,v,x^k\right)dx^i dx^j\,.
\end{equation}
In our choice of the coordinate system, we have
\begin{equation}
\begin{split}
&\delta g_{rr}=\delta g_{rv}=\delta g_{ri}=0\,,\quad\quad \delta g_{vv}\Big|_{r=0}=\delta g_{vi}\Big|_{r=0}=0\,, \quad \text{and,}\quad \delta g_{ij}=\delta h_{ij}\,.
\end{split}
\end{equation}
Let's compute different components of 
\begin{equation}
D_\sigma\delta g_{\nu\rho}=\partial_\sigma\delta g_{\nu\rho}-\Gamma^\beta_{\sigma\nu}\delta g_{\beta \rho}-\Gamma^\beta_{\sigma\rho}\delta g_{\nu \beta}\,.
\end{equation}
The non-vanishing components of $D_\sigma \delta g_{\nu\rho}$ are
\begin{equation}
\begin{split}
D_r \delta g_{vi}&=\delta \omega_i-\frac{1}{2}\omega^j \delta h_{ij}\,,\\
D_r\delta g_{ij}&=2\,\delta \bar{K}_{ij}-\bar{K}^\ell_i \delta h_{\ell j}-\bar{K}^\ell_j\delta h_{i\ell}\,,\\
D_v\delta g_{ri}&=-\frac{1}{2}\omega^j\delta h_{ji}\,,\\
D_v\delta g_{ij}&=2\,\delta K_{ij}-K^\ell_i\delta h_{\ell j}-K^\ell_j\delta h_{i\ell}\,,\\
D_i\delta g_{r j}&=-\bar{K}^\ell_i\delta h_{\ell j}\,,\\
D_i\delta g_{v j}&=-K^\ell_i\delta h_{\ell j}\,,\\
D_i\delta g_{jk}&=\nabla_i \delta h_{jk}\,.
\end{split}
\end{equation}
All other components of $D_\sigma \delta g_{\nu\rho}$ vanish:
\begin{equation}
\begin{split}
&D_r\delta g_{rr}=D_r\delta g_{rv}=D_r\delta g_{ri}=D_r\delta g_{vv}=D_v\delta g_{rr}=D_v\delta g_{rv}=0\,,\\
&D_v\delta g_{vv}=D_v\delta g_{vi}=D_i\delta g_{rr}=D_i\delta g_{rv}=D_i\delta g_{vv}=0\,.
\end{split}
\end{equation}
Therefore, $r$-component of the first term in \eqref{eq:expThm} becomes
\begin{equation}
\begin{split}
R^{r\rho\nu\sigma}D_\sigma\delta g_{\nu\rho}&=R^{rivr}D_r\delta g_{vi}+R^{rvir}D_r\delta g_{iv}+R^{rjir}D_r\delta g_{ij}+R^{rirv}D_v\delta g_{ri}+R^{rjiv}D_v\delta g_{ij}+R^{rjri}D_i\delta g_{rj}\\
&+R^{rjvi}D_i\delta g_{vj}+R^{rvji}D_i\delta g_{jv}+R^{rkji}D_i\delta g_{jk}\\
&=R^{rivr}\Big(D_r\delta g_{vi}+D_r\delta g_{iv}-D_v\delta g_{ri}\Big)+R^{rjir}\Big(D_r\delta g_{ij}-D_i\delta g_{rj}\Big)\\
&+R^{rjiv}\Big(D_v\delta g_{ij}-D_i\delta g_{vj}\Big)+R^{rvji} D_i\delta g_{jv}+R^{rkji}D_i\delta g_{jk}\,.
\end{split}
\end{equation}
As we want the above answer up to ${\cal O}(\delta g)$, we have to compute the Riemann tensor only on the {\it background metric}. Now the inverse metric components on $r=0$ are
\begin{equation}
g^{rv}=1,\quad g^{rr}=g^{ri}=g^{vv}=g^{vi}=0,\quad \text{and,}\quad g^{ij}=h^{ij}. 
\end{equation}
Now, using Appendix C of \cite{Bhattacharya:2019qal} for the components of Riemann tensor, we get
\begin{equation}
\begin{split}
R^{rivr}&=h^{ij}R_{vjrv}=-\frac{1}{2}h^{ij}\Big(\partial_v \omega_j+\omega^\ell K_{j\ell}\Big)=0\,,\\
R^{rjir}&=h^{j\ell}h^{i m}R_{v\ell m v}=h^{j\ell}h^{i m}\Big(\partial_v K_{\ell m}-K_{\ell n}K^n_m\Big)=0\,,\\
R^{rkji}&=h^{km}h^{jn}h^{i\ell}R_{vmn\ell}=h^{km}h^{jn}h^{i\ell}\Big(\nabla_j K_{ik}-\nabla_i K_{jk}-\frac{1}{2}\omega_i K_{jk}+\frac{1}{2}\omega_j K_{ik}\Big)=0\,.
\end{split}
\end{equation}
Therefore, we get
\begin{equation}
\begin{split}
R^{r\rho\nu\sigma}D_\sigma \delta g_{\nu\rho}&=R^{rjiv}(D_v\delta g_{ij}-D_i\delta g_{vj})+R^{rvji}D_i\delta g_{jv}\\
&=R^{rjiv}\Big(2\,\delta K_{ij}-\cancel{K_i^\ell\delta h_{\ell j}}-K^\ell_j\delta h_{i\ell}+\cancel{K^\ell_i\delta h_{\ell j}}\Big)+R^{rvji}(-K^\ell_i\delta h_{\ell j})\\
&=R^{rjiv}2\,\delta K_{ij}+{\cal O}(\delta g)^2\,.
\end{split}
\end{equation}
Now, let's compute the $r$-component of the second term in \eqref{eq:expThm}
\begin{equation}
(D_\sigma R^{r\nu\rho\sigma})\delta g_{\nu\rho}=(D_\sigma R^{rij\sigma})\delta h_{ij}\,.
\end{equation}
Now, $D_\sigma R^{rij\sigma}$ has boost weight $+1$ therefore, vanishes on the background.\\
So, finally we get
\begin{equation}
\begin{split}
\Theta^r&=8\,R^{rjiv}\delta K_{ij}\,.
\end{split}
\end{equation}

\subsection*{Computation of $Q^{rv}$}
Now, let's compute $Q^{rv}$ (see Table 5 of \cite{Bhattacharyya:2021jhr})
\begin{equation}\label{eq:expqrv}
Q^{rv}=4 R^{rv\alpha\sigma}D_\sigma \xi_\alpha-8(D_\sigma R^{rv\alpha\sigma})\xi_\alpha\,.
\end{equation}
Let's consider the first term in \eqref{eq:expqrv}
\begin{equation}
R^{rv\alpha\sigma}D_\sigma \xi_\alpha\,.
\end{equation}
Now, $\xi_\mu$ has the following expression
\begin{equation}
\xi_r=v,\quad \xi_v=-r-r^2 v X,\quad \xi_i=rv\omega_i\,.
\end{equation}
Different components of $D_\sigma\xi_\alpha$ are
\begin{equation}\label{eq:Drxi23}
\begin{split}
&D_r\xi_r=0,\quad D_r\xi_v=-1,\quad D_r\xi_i=\frac{1}{2}v\,\omega_i,\quad D_v\xi_r=1,\quad D_v\xi_v=0,\quad D_v\xi_i=0,\quad D_i\xi_r=-\frac{1}{2}v\,\omega_i,\\
& D_i\xi_v=0,\quad D_i\xi_j=v\,K_{ij}\,.
\end{split}
\end{equation}
Therefore we get
\begin{equation}
\begin{split}
R^{rv\alpha\sigma}D_\sigma \xi_\alpha&=R^{rvrv}D_v\xi_r+R^{rvri}D_i \xi_r+R^{rvvr}D_r \xi_v+R^{rvir}D_r \xi_i+R^{rvij}D_j\xi_i\\
&=2\,R^{rvrv}+R^{rvri}\Big(-\frac{1}{2}v\,\omega_i\Big)+R^{rvir}\frac{1}{2}v\,\omega_i+\cancel{R^{rvij}vK_{ij}}\\
&=2\,R^{rvrv}-v\,\omega_i R^{rvri}\,.
\end{split}
\end{equation}
Now, let's compute the second term in \eqref{eq:expqrv}
\begin{equation}
\xi_\alpha(D_\sigma R^{rv\alpha\sigma})=\xi_\alpha\Big(\partial_\sigma R^{rv\alpha\sigma}+\Gamma^r_{\sigma\beta}R^{\beta v\alpha\sigma}+\Gamma^v_{\sigma\beta}R^{r\beta\alpha\sigma}+\Gamma^\alpha_{\sigma\beta}R^{rv\beta\sigma}+\Gamma^\sigma_{\sigma\beta}R^{rv\alpha\beta}\Big)\,.
\end{equation}
On the $r=0$ hypersurface, $\xi_r=v\,,\,\, \xi_v=0\,,\,\,\xi_i=0$
\begin{equation}
\begin{split}
\xi_\alpha(D_\sigma R^{rv\alpha\sigma})&=v\Big(\partial_\sigma R^{rvr\sigma}+\Gamma^r_{\sigma\beta}R^{\beta vr\sigma}+\Gamma^v_{\sigma\beta}R^{r\beta r\sigma}+\Gamma^r_{\sigma\beta}R^{rv\beta\sigma}+\Gamma^\sigma_{\sigma\beta}R^{rvr\beta}\Big)\\
&=v\Big(\partial_v R^{rvrv}+\partial_i R^{rvri}+\frac{1}{2}\omega_i R^{rvri}-K_{ij}R^{jvri}-\omega_i R^{rvri}-\bar{K}_{ij}R^{rjri}+K R^{rvrv}+\hat{\Gamma}^i_{ij}R^{rvrj}\Big)\\
&=v\Big(\partial_v R^{rvrv}+\nabla_i R^{rvri}-\frac{1}{2}\omega_i R^{rvri}-K_{ij}R^{jvri}-\bar{K}_{ij}R^{rjri}+K R^{rvrv}\Big)\,.
\end{split}
\end{equation}
Therefore, finally we get
\begin{equation}
\begin{split}
Q^{rv}&=4 R^{rv\alpha\sigma}D_\sigma \xi_\alpha-8(D_\sigma R^{rv\alpha\sigma})\xi_\alpha\\
&=4(2\,R^{rvrv}-v\,\omega_i R^{rvri})-8v\Big(\partial_v R^{rvrv}+\nabla_i R^{rvri}-\frac{1}{2}\omega_i R^{rvri}-K_{ij}R^{jvri}-\bar{K}_{ij}R^{rjri}+K R^{rvrv}\Big)\\
&=8\,R^{rvrv}-8v\Big((\partial_v+K) R^{rvrv}+\nabla_i R^{rvri}-K_{ij}R^{jvri}-\bar{K}_{ij}R^{rjri}\Big)\,.
\end{split}
\end{equation}
\subsection*{Computation of $Q^{ri}$}
Now, let's compute $Q^{ri}$ (see Table 5 of \cite{Bhattacharyya:2021jhr})
\begin{equation}\label{eq:Qmnu}
Q^{ri}=4 R^{ri\alpha\sigma}D_\sigma \xi_\alpha-8(D_\sigma R^{ri\alpha\sigma})\xi_\alpha\,.
\end{equation}
Let's first consider the first term in \eqref{eq:Qmnu}
\begin{equation}
\begin{split}
R^{ri\alpha\sigma}D_\sigma\xi_\alpha&=R^{rirv}+R^{rirj}\Big(-\frac{1}{2}v\,\omega_j\Big)+R^{rivr}(-1)+R^{rijr}\Big(\frac{1}{2}v\,\omega_j\Big)+R^{rij\ell}v\,K_{\ell j}\\
&=2\, R^{rirv}-v\,\omega_j R^{rirj}+\cancel{v\,K_{\ell j}R^{rij\ell}}\,,
\end{split}
\end{equation}
where, we have used expressions of $D_\sigma\xi_\alpha$ as given in eqn.\,\eqref{eq:Drxi23}. Now, let's consider the term
\begin{equation}
\begin{split}
\xi_\alpha D_\sigma R^{ri\alpha\sigma}&=\xi_\alpha\Big(\partial_\sigma R^{ri\alpha\sigma}+\Gamma^r_{\sigma\beta}R^{\beta i\alpha\sigma}+\Gamma^i_{\sigma\beta}R^{r\beta\alpha\sigma}+\Gamma^\alpha_{\sigma\beta}R^{ri\beta\sigma}+\Gamma^\sigma_{\sigma\beta}R^{ri\alpha\beta}\Big)\\
&=v\Big(\partial_\sigma R^{rir\sigma}+\Gamma^r_{\sigma\beta}R^{\beta ir\sigma}+\Gamma^i_{\sigma\beta}R^{r\beta r\sigma}+\Gamma^r_{\sigma\beta}R^{ri\beta\sigma}+\Gamma^\sigma_{\sigma\beta}R^{rir\beta}\Big)\,.
\end{split}
\end{equation}
Now, we will quote the answers for different terms in the above equation
\begin{equation}
\begin{split}
\partial_\sigma R^{rir\sigma}&=\partial_v R^{rirv}+\partial_j R^{rirj}\,,\\
\Gamma^r_{\sigma\beta}R^{\beta ir\sigma}&=\frac{1}{2}\omega_j R^{rirj}-K_{j\ell}R^{\ell irj}\,,\\
\Gamma^i_{\sigma\beta}R^{r\beta r\sigma}&=2\, K^i_j R^{rjrv}+\hat{\Gamma}^i_{j\ell}R^{r\ell r j}\,,\\
\Gamma^r_{\sigma\beta}R^{ri\beta\sigma}&=0\,,\\
\Gamma^\sigma_{\sigma\beta}R^{rir\beta}&=K R^{rirv}+\hat{\Gamma}^\ell_{\ell j}R^{r i r j}\,.
\end{split}
\end{equation}
Using the above expressions, we get
\begin{equation}
\begin{split}
Q^{ri}&=4(2\, R^{rirv}-v\,\omega_j R^{rirj})\\
&-8v\Big[\partial_v R^{rirv}+\partial_j R^{rirj}+\frac{1}{2}\omega_j R^{rirj}-K_{j\ell}R^{\ell irj}+2\, K^i_j R^{rjrv}+\hat{\Gamma}^i_{j\ell}R^{r\ell r j}+K R^{rirv}+\hat{\Gamma}^\ell_{\ell j}R^{r i r j}\Big]\\
&=8\,R^{rirv}-8v\,\partial_v R^{rirv}-8v\, K R^{rirv}-8v\,\nabla_j R^{rirj}-8v\,\omega_j R^{rirj}+8v\, K_{j\ell}R^{\ell irj}-16v\, K^i_j R^{rjrv}\\
&=8R^{rirv}-8v\,\partial_v R^{rirv}-8v\,\nabla_j R^{rirj}-8v\,\omega_j\,R^{rirj}+\mathcal{O}(\epsilon^2)\,.
\end{split}
\end{equation}

\section{A toy example to see the mechanism of mismatch at non-linear order}\label{App:Toy}
In this appendix, we demonstrate the mechanism underlying the mismatch between the two sides of \eqref{c0},
\begin{equation}\label{eq:dynwal}
S_{\rm dyn} = (1- v\partial_v)S_{\rm Wall}\,,
\end{equation}
at the non-linear order in perturbation theory through a simple example. As neither $S_{\rm Wall}$ nor $S_{\rm dyn}$ constitutes a valid notion of black hole entropy beyond linear order, there is no expectation that \eqref{eq:dynwal} should continue to hold at the non-linear level. Nevertheless, since constructing a consistent entropy functional beyond linear order is an important open problem, it is worthwhile to investigate the non-linear terms present in the naive definition of $S_{\rm Wall}$ and $S_{\rm dyn}$ and to identify precisely the origin of their mismatch.\\\\
We are working in our horizon adapted coordinate system \eqref{metric1}
\begin{equation}\label{eq:metap}
\begin{split}
ds^2 = 2\,dv\,dr - r^2 X\left(r,v,x^k\right)\,dv^2 + 2 r\,\omega_i\left(r,v,x^k\right)dv\,dx^i + h_{ij}\left(r,v,x^k\right)dx^i\,dx^j\,.
\end{split}
\end{equation}
$S_{\rm Wall}$ can be decomposed as \eqref{decomp}
\begin{equation}\label{Swall}
\begin{split}
\quad S_{\rm Wall} = S_{\rm Wall}^{\rm zero} + S_{\rm Wall}^{\rm rest}\,.
\end{split}
\end{equation}
 As a simple illustrative example, let us consider
\begin{equation}
\begin{split}
S^{\rm zero}_{\rm Wall}&= f(h_{ij}),~~S^{\rm rest}_{\rm Wall}= \alpha \left(\partial_v h_{ij}\right)\left(\partial_r h^{ij}\right)\,.
\end{split}
\end{equation}
On the other hand, $S_{\rm dyn}$ is defined as \eqref{defdyn}
\begin{equation}
\begin{split}
S_{\rm dyn}&=2\pi\int_{\Sigma_v}d^{D-2}x\,\sqrt{h}\,\Big[Q^{rv}(g,\xi) - v\,{\cal B}^r(g,\xi)\Big]\\
&\equiv S_{Q^{rv}}+v\, S_{\mathcal{B}^r}\\
&\equiv S^{\rm zero}_{Q^{rv}}+S^{\rm JKM}_{Q^{rv}}+v\, S_{\mathcal{B}^r}\,.
\end{split}
\end{equation}
According to the decomposition of \eqref{decomp}, $S_{\rm dyn}^{\rm zero}=S^{\rm zero}_{Q^{rv}}$ and $S_{\rm dyn}^{\rm rest}=S^{\rm JKM}_{Q^{rv}}+v\, S_{\mathcal{B}^r}$.\\\\
$S_{\rm Wall}$ and $S_{\rm dyn}$ satisfy the following conditions:
\begin{enumerate}
\item $S_{\mathcal{B}^r}$ is a boost weight one term. Note that, $v\,S_{\mathcal{B}^r}$ is also a JKM term.
\item $S_{\rm dyn} = S_{\rm Wall}$ when the metric \eqref{eq:metap} is stationary. This will ensure that the difference between $S_{\rm dyn}$ and $S_{\rm Wall}$ (if it exists) must be of JKM type.
\item If we substitute $g_{ab} = \bar g_{ab}\left(rv,x^i\right) + \delta g_{ab}\left(r,v,x^i\right)$ then 
\begin{equation}
\begin{split}
vE_{vv}&=-v \partial_v^2S_{\rm Wall}\\
&=\partial_v\left[(1-v\partial_v)S_{\rm Wall}\right]=\partial_v S_{\rm dyn}\,,
\end{split}
\end{equation} 
where, we have neglected terms with more than one positive boost weight factors.
\end{enumerate}
In this toy example we would like to construct $S_{\rm dyn}$ satisfying the above conditions and we would like to see given an expression for $S_{\rm Wall}$,  how much freedom $S_{\rm dyn}$ has, if we demand equality only upto  linear order in $\delta g$. Let us first note that
\begin{equation}\label{evv}
\begin{split}
vE_{vv} =~& - v\partial_v^2S_{\rm Wall} + \text{terms with at least two positive boost weight factors}\\
=~& -v\partial_v\left[\left(\partial f\over \partial h_{ij}\right)\partial_v h_{ij}\right] - v\alpha \left[2(\partial_v\partial_r h^{ij})~(\partial_v^2 h_{ij}) + (\partial_r h^{ij})~(\partial_v^3 h_{ij})\right]\\
&+\text{terms with at least two positive boost weight factors}\,.
\end{split}
\end{equation}
We shall start by constructing the most general structure of $S_{\rm dyn}$, keeping in mind that it has to generate the above terms in $E_{vv}$. Note that 
\begin{enumerate}
\item $S_{\rm dyn}$ can have maximum one $\partial_r$ operator.
\item $S_{Q^{rv}}^{\rm zero} = S_{\rm Wall}^{\rm zero}$ since $S_{\rm Wall}$ and $S_{\rm dyn}$ must match in equilibrium.
\item $S_{Q^{rv}}^{\rm JKM}$ is a JKM type term with boost weight zero. Since we can have only one $\partial_r$ operator in this case, the only term possible here is $(\partial_r h^{ij})(\partial_v h_{ij})$.
\item $\mathcal{B}^r$ is a boost-weight one term, so the possible candidates (restricting to the fact that there could be only one $\partial_r$) are $\partial_v h_{ij},\,\,(\partial_r h^{ij})(\partial_v^2 h_{ij}),\,\,(\partial_r\partial_v h^{ij})(\partial_v h_{ij})$\,.
\end{enumerate}
So, the most general form of $S_{\rm dyn}$ in this case
\begin{equation}\label{sdyne}
\begin{split}
S_{\rm dyn} &=S^{\rm zero}_{Q^{rv}}+S^{\rm JKM}_{Q^{rv}}+v\, S_{\mathcal{B}^r}\\
S^{\rm zero}_{Q^{rv}} &= f(h_{ij}),\qquad S^{\rm JKM}_{Q^{rv}}=\beta ~(\partial_r h^{ij})(\partial_v h_{ij})\\
S_{\mathcal{B}^r} &=\gamma_1 \left(\partial f\over \partial h_{ij}\right)\partial_v h_{ij} +\gamma_2  (\partial_r h^{ij})(\partial_v^2 h_{ij}) +\gamma_3 (\partial_r\partial_v h^{ij})(\partial_v h_{ij})\,.
\end{split}
\end{equation}
Now, we shall compute $\partial_v S_{\rm dyn}$.
\begin{equation}\label{delv}
\begin{split}
vE_{vv}&=\partial_v S_{dyn}\\
& =(1+\gamma_1) \left(\partial f\over \partial h_{ij}\right)\partial_v h_{ij} +\gamma_1\,v\partial_v \left[\left(\partial f\over \partial h_{ij}\right)\partial_v h_{ij}\right]+\beta(\partial_v\partial_r h^{ij})(\partial_v h_{ij})+\beta(\partial_r h^{ij})(\partial_v^2 h_{ij})\\
& + \gamma_2  (\partial_r h^{ij})(\partial_v^2 h_{ij}) +\gamma_3 (\partial_r\partial_v h^{ij})(\partial_v h_{ij})+ \gamma_2\,v (\partial_r h^{ij})(\partial_v^3 h_{ij}) + v(\gamma_2+\gamma_3)(\partial_r\partial_v h^{ij})(\partial_v^2 h_{ij})\,.
\end{split}
\end{equation}
Equating \eqref{delv} with \eqref{evv} and treating $\partial_v h_{ij}$, $\partial_v^2 h_{ij}$ and $\partial_v^3 h_{ij}$ as independent, we get the following constraints for the respective coefficients.
\begin{equation}\label{retr}
\begin{split}
&\gamma_1=-1\,,\\
\text{Coeff of $\partial_v h$}:~&(\beta +\gamma_3)~ \partial_r\partial_v h^{ij} =0,~~\Rightarrow~~\gamma_3 = -\beta\,,\\
\text{Coeff of $\partial_v^2 h$}:~&(\beta + \gamma_2)(\partial_r h_{ij}) + v(\gamma_2+\gamma_3+2\alpha)(\partial_r\partial_v h^{ij}) =0\,,\\
\text{Coeff of $\partial_v^3 h$}:~&v(\gamma_2 +\alpha) (\partial_r h^{ij}) =0,~~\Rightarrow~~\gamma_2 = -\alpha\,.\\
\end{split}
\end{equation}
%It is the coefficient of $\partial_v^2 h_{ij}$ that requires attention. If we want $(1-v\partial_v)S_{Wall}$ and $S_{\rm dyn}$ to match upto all orders in $\delta h$, then $(\partial_r h_{ij})$ and $(\partial_r\partial_v h^{ij})$ should be treated as independent structures, leading to two constraints between the coefficients
%$$\beta +a_1 =0,~~a_1 +a_2 + 2\alpha =0$$
%In this case the only consistent solution is 
%$$a_1 =a_2 =-\alpha,~~\beta = \alpha,~~\Rightarrow~~(1-v\partial_v)S_{Wall} = S_{dyn}~~\text{Exactly to all orders}$$
Now, the claim is that $S_{\rm dyn}$ and $(1-v\partial_v)S_{\rm Wall}$ should match upto the linear order in $\delta g$, which means in \eqref{retr} all $g_{ab}(r,v,x^i)$ should be replaced by $\bar g_{ab}(rv,x^i)$, satisfying
$$(1-v\partial_v)\partial_r\bar g_{ab}(rv,x^i) =0\,.$$
Now, the coefficient of $\partial_v^2h_{ij}$ leads to the following single constraint between the coefficients.
$$\beta +\gamma_3 +2\gamma_2 +2\alpha =0\,.$$
This is automatically true as soon as we ensure the vanishing of the coefficients of $\partial_v h_{ij}$ and  $\partial_v^2 h_{ij}$.\\
So, if we demand equality only upto linear order in $\delta g$,  the following could be a valid candidate for $S_{\rm dyn}$ for any constant $\alpha$ and $\beta$
\begin{equation}
\begin{split}
S_{\rm dyn} &=S^{\rm zero}_{Q^{rv}}+S^{\rm JKM}_{Q^{rv}}+v\, S_{\mathcal{B}^r}\,,\\
S^{\rm zero}_{Q^{rv}} &= f(h_{ij}),\qquad S^{\rm JKM}_{Q^{rv}}=\beta ~(\partial_r h^{ij})(\partial_v h_{ij})\,,\\
S_{\mathcal{B}^r} &=-\left(\partial f\over \partial h_{ij}\right)\partial_v h_{ij} -\alpha  (\partial_r h^{ij})(\partial_v^2 h_{ij}) -\beta (\partial_r\partial_v h^{ij})(\partial_v h_{ij})\,.
\end{split}
\end{equation}
Now, it is easy see the difference between $S_{\rm dyn}$ and $S_{\rm Wall}$
$$(1 - v\partial_v)S_{\rm Wall} -S_{\rm dyn} = (\alpha-\beta)\left[(1-v\partial_v)(\partial_r h^{ij})\right](\partial_v h_{ij})\,.$$
Note that the factor $(1-v\partial_v)(\partial_r h^{ij})$ vanishes if we assume $h^{ij}$ is stationary {\it i.e.}, $h^{ij}(r,v,x^k)\rightarrow~\bar h^{ij}(rv,x^k)$.

\bibliographystyle{JHEP}
\bibliography{WaldvsWall}

\end{document}